\documentclass[%
reprint,
 amsmath,amssymb,
 aps,
]{revtex4-1}

\usepackage[T1]{fontenc} 
\usepackage{graphicx}
\usepackage{dcolumn}
\usepackage{bm}
\usepackage{mathtools}
\usepackage{breqn}
\usepackage{physics}
\usepackage{caption}
\usepackage{subcaption}
\usepackage{subcaption}
\DeclareCaptionLabelSeparator{none}{ }
\usepackage[section]{placeins}
\usepackage{mathrsfs}
\begin{document}


\title{Bianchi IX and VIII Quantum Cosmology with a Cosmological Constant, Aligned Electromagnetic Field, and Scalar Field }

\author{Daniel Berkowitz}
 \altaffiliation{Physics Department, Yale University.\\ daniel.berkowitz@yale.edu \\ This work is in memory of my parents, Susan Orchan Berkowitz, and Jonathan Mark Berkowitz}

\date{\today}

\begin{abstract}
We investigate the quantum cosmologies of the Bianchi IX and VIII models when a cosmological constant, aligned electromagnetic field and free scalar field are present. The conserved quantity $p_{\phi}$ associated with our free scalar field results in $\phi$ classically being a quantity which monotonically increases with respect to time, thus allowing it to fulfill the role of an 'emergent' internal clock for our constrained quantum systems. We embark on this investigation to better understand how matter sources can affect general anisotropic quantum cosmologies. To aid us we use the Euclidean-signature semi classical method to obtain our wave functions and analyze them. In addition we study briefly the quantum Taub models when an $e^{ \phi}$ potential and aligned electromagnetic field are present. One of the interesting things we found was that our aligned electromagnetic field, depending on how strong it is, can create or destroy geometric states in our 'excited' state wave functions that quantum Bianchi IX universes can tunnel in and out of. This creation of a state is somewhat similar to how non-commutativity in the minisuperspace variables can cause new quantum states to emerge in the quantum Kantowski-Sachs and Bianchi I models. Our results further show the utility of the Euclidean-signature semi classical method towards tackling Lorentzian signature problems without having to invoke a Wick rotation. This feature of not needing to apply a Wick rotation makes this method potentially very useful for tackling a variety of problems in bosonic relativistic field theory and quantum gravity.

\end{abstract}

\pacs{Valid PACS appear here}
\maketitle


\section{\label{sec:level1}INTRODUCTION}
Even though on large scales our universe is incredibly isotropic and homogeneous it is highly likely that our early universe possessed a considerable degree of anisotropy and inhomogeneity which originated from quantum fluctuations within particle fields whose size was comparable to the primordial cosmological horizon. Thus it is useful to study anistropic or inhomogeneous classical/quantum cosmologies so we can better understand what our universe could have been like when it was extremely young. To accomplish this we will study the quantum Bianchi IX, VIII and Taub models with matter sources in order to obtain an idea of what possible effects those matter sources could have induced in our early universe. In the future it would be worthwhile to include perturbative inhomogeneities in either the electric/magnetic or scalar field. 

The wave functions we will obtain are easiest to interpret in a qualitative manner when we allow $\Lambda < 0$. Choosing a negative cosmological constant does not necessarily make our solutions non-physical. Recently there has been some interesting work\cite{biswas2009inflation,hartle2012accelerated,mithani2013inflation, hartle2014quantum,visinelli2019revisiting} done in studying inflation with a negative cosmological constant and the connection  between asymptotically AdS wave functions to classical cosmological histories which exhibit phenomenology that one expects from universes with a positive cosmological constant. From a theoretical point of view it is worthwhile to study quantum cosmologies that possess a negative cosmological constant. 

In addition we will only consider an electromagnetic field in which the electric and magnetic components of it are parallel with each other and are both non-vanishing. The case when only a primordial magnetic field is present can be realized in the quantum Bianchi I\cite{ryan1982bianchi} models which can also be thoroughly analyzed with the method we will employ in this paper.

Despite these restrictions the wave functions we obtain exhibit fascinating properties. This is especially true in regards to how our aligned electromagnetic field affects the 'excited' states of
our quantum models which we computed using the Euclidean-signature semi classical method. As a result they provide further incentives to study wave functions of the universe derived from more general electric/magnetic field configurations.

Beyond theoretical considerations, new evidence\cite{neronov2010evidence,tavecchio2010intergalactic} for the existence of a femto Gauss strength intergalactic magnetic field has been uncovered by observing gamma rays. This provides further reasons to continue \cite{kamenshchik1993fermions,esposito1995relativistic,kobayashi2019early,jimenez2009cosmological,louko1988quantum,karagiorgos2018quantum,pavvsivc2012wheeler} studying electric/magnetic fields through the lens of quantum cosmology. Through studying the effects of electric/magnetic fields on quantum universes we can potentially better understand how seeds of anisotropy developed in our early universe which we can observe\cite{bennett2013nine,hinshaw2013nine,ade2016planck} today in the CMB. Recent work\cite{kahniashvili2001cmb,paoletti2011cmb,miyamoto2014cmb,hortua2017reduced} has been conducted in trying to determine what signatures a primordial magnetic field would induce on the various computable spectrums derived from the CMB. By studying what effects aligned electromagnetic fields induce on both 'ground' state and 'excited' state wave functions of quantum cosmological models we can  contribute to the theoretical portion of that task.

The Bianchi IX models possess a rich history and have been studied in a plethora of contexts \cite{ryan1975relativistic,graham1991supersymmetric,graham1993anisotropic,graham1993supersymmetric,bene1994supersymmetric,graham1994hartle,d1993quantization,csordas1995supersymmetric} Investigations into them began with the works of Misner, Ryan, and Belinskii, Khalatnikov and Lifshitz\cite{misner1969mixmaster,ryan1972oscillatory,belinskii1970oscillatory,belinskii1978asymptotically}. What partly made these models so appealing was that they shared the 3-sphere spatial topology of the k=1 FLRW models, which were widely believed to be a good approximation for our physical universe until more precise cosmological observations \cite{de2002multiple} showed otherwise. In addition the equations which govern the dynamics of the Bianchi IX mini-superspace variables, which we will take to be the Misner variables $(\alpha,\beta_+,\beta_-)$ \cite{misner1969mixmaster,misner1969quantum,barrow1982chaotic,chernoff1983chaos,cornish1997mixmaster} admit chaotic solutions. It was originally thought that the chaos present near its singularity, which took the form of an erratic sequence of Kasner contractions and expansions could explain why we observe the universe to be mostly homogeneous, hence the origin of the name Mixmaster\cite{misner1969mixmaster}. 

The quantum diagonalized Bianchi IX models were first studied by Misner \cite{misner1969quantum}, and later by Moncrief-Ryan \cite{moncrief1991amplitude}; among others in a plethora of different contexts, such as supersymmetric quantum cosmology\cite{graham1992supersymmetric,macias1993supersymmetric}. They were later revisited in the 2010s by Bae and Moncrief\cite{bae2015mixmaster,moncrief2014euclidean}, when they applied the Euclidean-signature semi classical method\cite{marini2019euclidean,moncrief2014euclidean} to prove for the wormhole case that a smooth and globally defined asymptotic solution exists for arbitrary Hartle-Hawking\cite{hartle1983wave} ordering parameter. They also studied the 'excited' states of the Bianchi IX models. The quantum cosmology of Bianchi IX models with a cosmological constant was previously investigated by employing Chern-Simons solutions in Ashtekar’s variables \cite{paternoga1996exact,graham1996physical}.

The Bianchi VIII quantum/classical models have not been as thoroughly studied as the Bianchi IX models. However, there has been some interesting investigations\cite{banerjee1984spatially,lorenz1980exact,obregon1996psi} into them and exact solutions of them have been obtained in both classical and quantum regimes. 

The classical Taub models \cite{taub1951empty} and their extension, the Taub-NUT models have a rich history of their own, which includes being used to model the space-time around a black hole \cite{newman1963empty,kerner2006tunnelling}. Recently work has been conducted in finding new solutions to the symmetry reduced Wheeler DeWitt equation of the LRS Bianchi IX models \cite{karagiorgos2019quantum} using Killing vectors and tensors. Their quantum cosmology has also been investigated within the context of the generalized uncertainty principle \cite{battisti2008quantum} and using the WKB approximation\cite{cascioli2019wkb,de2020dynamics}.  In addition the author proved\cite{berkowitz2020towards} the existence of a countably infinite number of 'excited' states for the quantum Taub models when a cosmological constant is present and is currently working on proving their existence when both a cosmological constant and aligned electromagnetic field are present. We will further expand upon what was previously done by obtaining a leading order or closed form solution depending on the operator ordering of the Wheeler DeWitt equation for the quantum Taub models when an exponential $e^{\phi}$ scalar field potential is present.

This paper will have the following structure. First we will introduce our models and discuss the problem of time associated with them. Afterwards we will show explicitly show how we quantize the electromagnetic field in our models to form their Wheeler Dewitt equations. Then we will introduce the Euclidean-signature semi classical method. From there we will study the Bianchi IX models in the semi classical limit when a cosmological constant, aligned electromagnetic field and stiff matter are present. The semi-classical wave functions we obtain in closed form for the case when a negative cosmological constant is present behaves qualitatively in a similar manner to the wave functions reported in \cite{paternoga1996exact}. This supports the idea that our semi-classical wave functions reasonably capture the effects of our matter sources. In addition, the fact that we were able to find closed form solutions to the Euclidean-signature Hamilton Jacobi equation which can be used to construct semi-classical solutions to the Lorentzian signature symmetry reduced WDW equation without needing to use a Wick rotation further shows the promise of this method to be a useful alternative\cite{moncrief2014euclidean} to Euclidean path integrals\cite{gibbons1979positive,gibbons1993euclidean} for tackling problems in quantum gravity.  

Next we will study wave functions of the Bianchi IX models with our matter sources which have as their classical analogues trajectories in minisuperspace in which $\beta_{-}$ is fixed at zero. Next we will turn our attention to the Taub models and briefly analyze them when an exponential scalar field and aligned electromagnetic field potentials are present.  We will interpret our wave functions by their atheistic characteristics. For example we will assume, as was done in \cite{garcia2002noncommutative}, that each visible peak which is present for our wave functions represents a geometric state a quantum universe can tunnel in and out of. Finally we will present a few closed form solutions to the WDW equation for the Bianchi IX and VIII models that the author found\cite{berkowitz2019new} and give some concluding remarks. 

\section{\label{sec:level2}Our Models}
All Bianchi cosmological models can be represented by the following metric 

\begin{equation}\label{1}
\begin{aligned}
&d s^{2}=-N^{2}d t^{2}+\frac{L^{2}}{6\pi}e^{2 \alpha(t)}\left(e^{2 \beta(t)}\right)_{i j} \omega^{i} \omega^{j}\\& \left(e^{2 \beta(t)}\right)_{i j}=e^{2\beta_{+}(t)}\operatorname{diag}\left(e^{2\sqrt{3}\beta_{-}(t)}, e^{-2\sqrt{3}\beta_{-}(t)}, e^{-6 \beta_{+}\left(t\right)}\right).
\end{aligned}
\end{equation}
The $\omega^{i}$ terms are one forms defined on the spatial hypersurface of each Bianchi cosmology and obey $ d \omega^{i}=\frac{1}{2} C_{j k}^{i} \omega^{j} \wedge \omega^{k}$ where $C_{j k}^{i}$ are the structure
constants of the invariance Lie group associated with each particular class of Bianchi models. We choose to make our one forms $\omega^{i}$ dimensionless by introducing $L$ which has units of 'length' and sets a length scale for the spatial size of our cosmology. This can be seen because any shift in the scale factor $e^{\alpha(t)+\delta}$ where $\delta$ is a real number can be reabsorbed into $L^{2}$. Beyond inspecting (\ref{1}), it can be seen that $e^{\alpha(t)}$ acts as the scale factor for our models in these Misner variables by computing $\sqrt{-\operatorname{det} g}$ of the metric tensor (\ref{1}) expressed in orthonormal coordinates, yielding $e^{3\alpha(t)}$ which measures the total volume of our Bianchi IX space-time and the relative volume in a region for our Bianchi VIII space-time.

For the Bianchi IX and VIII models their one forms are 

\begin{equation}\label{2}
\begin{aligned}
\omega^{1} &=d x-k \sinh (k y) d z \\
\omega^{2} &=\cos (x) d y-\sin (x) \cosh (k y) d z \\
\omega^{3} &=\sin (x) d y+\cos (x) \cosh (k y) d z
\end{aligned}
\end{equation}
where k=1 is for the Bianchi VIII models and k=$\mathrm{i}$ is for the Bianchi IX models. 

Upon setting $c=1, G=\frac{1}{16\pi}$, and $L=1$ we can express the Einstein-Hilbert action with matter as 

\begin{equation}\label{3}
\begin{aligned}
&S_{E-H}=\int \sqrt{-\operatorname{det} g} \left(R -\Lambda+\frac{1}{2} g^{\mu \nu} \partial_{\mu} \phi \partial_{\nu} \phi\right) d^{4} x \\& + S_{matter}.
\end{aligned} 
\end{equation}
If we only concern ourselves with the gravitation and scalar field sector of the E-H action we can obtain the following action in terms of ADM variables

\begin{equation}\label{4}
\begin{aligned}
&\mathcal{S_{ADM}}=  N \sqrt{h}\Biggl(K_{i j} K^{i j }-K^{2}+R-\Lambda \\&+\frac{1}{2} g^{\mu \nu} \partial_{\mu} \phi \partial_{\nu} \phi\Biggr)d^{4} x ,
\end{aligned}
\end{equation}
where $K_{i j}$ are the components of the extrinsic curvature which measures the curvature induced on the Riemannian manifold equipped with spatial metric $h_{i j}$, from the higher dimensional space-time $g_{\mu \nu}$ it is embedded in, and $K$ is the trace of the extrinsic curvature. 

The kinetic term belonging to the Hamiltonian constraint which can be derived from (\ref{4}) is
\begin{equation}\label{5}
e^{-3 \alpha}\left(-p_{\alpha}^{2}+p_{+}^{2}+p_{-}^{2}+12p_{\phi}^{2}\right)
\end{equation}
and can be quantized as follows \begin{equation}\label{6}
\begin{array}{l}{-e^{-3 \alpha} p_{\alpha}^{2} \longrightarrow \frac{\hbar^{2}}{e^{(3-B) \alpha}} \frac{\partial}{\partial \alpha}\left(e^{-B \alpha} \frac{\partial}{\partial \alpha}\right)} \\ {e^{-3 \alpha} p_{+}^{2} \longrightarrow \frac{-\hbar^{2}}{e^{3 \alpha}} \frac{\partial^{2}}{\partial \beta_{+}^{2}}} \\ {e^{-3 \alpha} p_{-}^{2} \longrightarrow \frac{-\hbar^{2}}{e^{3 \alpha}} \frac{\partial^{2}}{\partial \beta_{-}^{2}}}\\ {e^{-3 \alpha} p_{\phi}^{2} \longrightarrow \frac{-\hbar^{2}}{e^{3 \alpha}} \frac{\partial^{2}}{\partial \phi^{2}}}\end{array}
\end{equation}

Using these quantized canonical momenta and the well known\cite{uggla1995classifying,obregon1996psi} contributions to the gravitational and matter components of the potentials for the Bianchi A models we can write down the Wheeler DeWitt equations that will aid us in studying quantum cosmologies of the Bianchi VIII, IX(\ref{7}), and Taub models(\ref{8}) where $\rho$ is a constant stiff matter term and $\lor$ is the logical "or" symbol, and $U_{+}$ is the Bianchi VIII potential while $U_{-}$ is the Bianchi IX potential

\begin{equation}\label{7}
\begin{aligned} 
& \square \Psi-B \frac{\partial \Psi}{\partial \alpha}+U \Psi=0 \\ &
U_{\pm}=\left(f\right) e^{6 \beta_+} \left(e^{6 \beta_+} \sinh ^2\left(2 \sqrt{3}
   \beta_-\right)\pm\cosh \left(2 \sqrt{3} \beta_-\right)\right) \\& +\frac{2 e^{6 a} \Lambda}{9 \pi }+U_{em}+\frac{1}{4}f+\rho \\&
   U_{em}= 2b^{2}e^{2\alpha-4\beta_+} \lor 2b^{2}e^{2 \left(\alpha\pm\sqrt{3}\beta_-+\beta_+\right)}  \\& 
   f=\frac{4}{3} e^{4 \alpha-8 \beta_+} ,
\end{aligned}
\end{equation}

\begin{equation}\label{8}
\begin{aligned} 
&\frac{\partial^2 \Psi}{\partial \alpha^2}-B\frac{\partial \Psi}{\partial \alpha}-\frac{\partial^2 \Psi}{\partial \beta_+^{2}}-12\frac{\partial^2 \Psi}{\partial \phi^{2}}+V\Psi=0 \\& V=\left( \frac{e^{4 \alpha-8 \beta_+}}{3} \left(1-4 e^{6 \beta_+} \right)+e^{6\alpha +\phi}\right)+2b^2e^{2\alpha+2\beta_{+}}.
\end{aligned}
\end{equation}
In (\ref{7}) $\square$ has signature $\left(+1,-1,-1,-1\right)$. Also our equations are written using the Hartle-Hawking\cite{hartle1983wave} operator ordering where $B$ is the Hartle-Hawking ordering parameter. We will derive the electromagnetic components of the potentials in the next section. In addition we will find solutions to the Wheeler DeWitt equation using a different operator ordering that we will introduce later. 

The Wheeler DeWitt equations (\ref{7} and \ref{8}) are the analogues to the time dependent Schr$\text{\" o}$dinger equations for our quantum cosmologies. Viewing the Wheeler DeWitt(WDW) equation as $\hat{\mathcal{H}}_{\perp} \Psi=0$, and trying to relate it to the conventional Schr$\text{\" o}$dinger equation results in the problem of time manifesting itself as
\begin{equation}\label{9}
i \hbar \frac{\partial \Psi}{\partial t}=N \hat{\mathcal{H}}_{\perp} \Psi=0,
\end{equation}
where $\frac{\partial \Psi}{\partial t}=0$. Due to the absence of the time derivative term of the Schr$\text{\" o}$dinger equation in the WDW equation, the construction of a unitary time evolution operator is not trivial, thus leading to the potential breakdown of a simple probabilistic interpretation of the wave function of the universe.

A Klein-Gordon current 
 \begin{equation}\label{10}
\mathcal{J}=\frac{i}{2}\left(\Psi^{*} \nabla \Psi-\Psi \nabla \Psi^{*}\right)
\end{equation}
can be defined \cite{vilenkin1989interpretation,mostafazadeh2004quantum}  which could be used to construct a probability density. It however, possesses unattractive features such as it vanishing when the wave function used to construct the current is purely real or imaginary and not always being positive definite. 

Besides the issue of constructing a probability density function, it appears the quantized Hamiltonian constraint admits only zero's as eigenvalues. This may lead one to the conclusion that all of the states which satisfy the WDW equation possess vanishing energy. This on the surface makes it impossible to distinguish between ground and excited states because all states seemingly have the same energy. This apparent obstacle to delineate 'ground' and 'excited' states can be overcome by examining the nuanced nature of the ADM formalism \cite{arnowitt1959dynamical}. When cast in the ADM formalism general relativity is a constrained theory with four Lagrange multipliers, the lapse and the three components of the shift. The constraint associated with the lapse is due to general relativity being invariant under reparameterization of the evolution parameter. Likewise the constraint associated with the shift is due to diffeomorphism invariance and is called the diffeomorphism constraint. The diffeomorphism constraint is due to the configuration space $h_ {ab} $ being too large to the point of it being physically redundant. To remedy this one can define a superspace \cite{fischer1970theory,giulini2009superspace}  where an equivalence class for $h_ {ab} $ is constructed such that two $h_ {ab} $ are in the same class if they can be carried into one another by a diffeomorphism. This shrinks the configuration space, allowing the diffeomorphism constraint to be satisfied. The same cannot be done for the reparameterization constraint \cite{wald1984general}. This explains why it wouldn't even make sense for a time derivative to be present because there is no unique "time" to use and partially explains the origins of the "problem of time". 

To get a better feel for what is going on, one can examine the vanishing Hamiltonian of a fully constrained system. One can formulate the Lagrangian of a free particle moving in one dimension , and introduce another configuration variable by defining the function t(T) where T is some arbitrary evolution parameter. If one were to treat both X(t(T)) and t(T) as configuration variables and formulate the system's Hamiltonian, they would notice that the Hamiltonian vanishes. Obviously the energy of a free one dimensional particle moving at a particular velocity cannot be zero. This is resolved by realizing that the dynamics of the system are now encoded in how X(t(T)) evolves with respect to t(T) where both are configuration variables. For an explicit demonstration of the above vanishing Hamiltonian construction, we refer the reader to \cite{rovelli2014covariant}. In other words, for these types of constrained systems the Hamiltonian no longer corresponds to the total energy. Thus the Hamiltonian constraint we quantized does not represent the total energy of a space-time in general relativity, and its vanishing eigenvalues do not mean that only states which possess zero energy are physically allowed. This allows leeway in defining 'ground' and 'excited' states in which features of ordinary quantum mechanics manifest as will be demonstrated in section 4.

A more in depth discussion in regards to how the Euclidean-signature semi classical method can be used to define 'ground' and 'excited' states despite them both being annihilated by the quantized Hamiltonian constraint can be found in \cite{moncrief2014euclidean}. To deal with the problem of time we will choose one of our variables $\left(\alpha,\beta_{+},\beta_{-},\phi\right)$ to act as an evolution parameter \cite{dewitt1967quantum}. Both $\alpha$ and $\phi$ are good candidates to be our "clock" for various reasons. The scale factor $\alpha$ would be an intuitive clock to use because it corresponds to size of our models and thus is fundamentally intrinsic to them. In addition if we look at (\ref{5}) we can rewrite it as 
\begin{equation}\label{11}
\begin{aligned} 
&e^{-3\alpha}p^{i}G_{i j}p^{j}\\&
\Vec{p}=\left(p_{\alpha},p_{+},p_{-},p_{\phi}\right)\\&
G_{i j}=\operatorname{diag}\left(-1,1,1,12\right),
\end{aligned}
\end{equation}
where the signature associated with $\alpha$ has the same sign as the time component of our metric (\ref{1}). A downside to using $\alpha$ as an internal clock though is that classically Bianchi IX universes experience a transition from an expansionary epoch to contractionary one which causes $\alpha$ to "tick" backwards. An alternative is to use our free scalar field $\phi$ as our internal clock as has been done in loop quantum cosmology\cite{ashtekar2006quantum,ashtekar2006bquantum,ashtekar2006cquantum}. Because $p_{\phi}$ is a conserved quantity for our Bianchi A models(excluding our Taub model with a $e^{\phi}$ scalar filed) classically $\phi$ always increases monotonically and thus in a quantum cosmological context emerges as a candidate for time. This is not the case though when we add our exponential potential term to the Taub models (\ref{8}). In this paper we will use both $\alpha$ and $\phi$ as our internal clocks depending on what points the author wishes to elucidate.

\section{\label{sec:level3}Electromagnetic Potentials For The Taub Models}

In this section we will compare two methods for obtaining the WDW equations (\ref{7}). The first method will be based on directly quantizing the class of classical Hamilitonians for Bianchi A models that was developed in \cite{waller1984bianchi}. This will lead to a semi-classical treatment of our electromagnetic degree of freedom and will be what we use in the following sections to analyze how matter sources affect our wave functions. However we will also do a full quantum treatment of the electromagnetic degree of freedom and compare the two approaches. We will assume all of our electric and magnetic fields are parallel to each other as was done in \cite{hughston1970homogeneous,waller1984bianchi}.

With this in mind our first task is to obtain solutions for Maxwell's equations in the space-time (\ref{1}) in terms of the Misner variables. In our calculations we will set $L=\sqrt{6\pi} \ell$ where $\ell$ is a quantity which equals unity and has some arbitrary unit of length. Starting from 
\begin{equation}\label{12}
\boldsymbol{A}=A_{0}dt+A_{1}\omega^{1}+A_{2}\omega^{3}+A_{3}\omega^{3},
\end{equation}
where $\omega^{0}=dt$,
and using the fact that $  d\omega^{i}=\frac{1}{2} C_{j k}^{i} \omega^{j} \wedge \omega^{k}$
to aide us in computing $\boldsymbol{F}=d\boldsymbol{A}= \frac{1}{2}F_{\mu v}\omega^{\mu} \wedge \omega^{v}$ results in the following expression for $F_{\mu v}$
\begin{equation}\label{13}
 F_{\mu v}=A_{v,\mu}-A_{ \mu,v}+A_{\alpha} C_{\mu v}^{\alpha}.
\end{equation}
In (\ref{13}) differentiation is done through a vector dual to our one forms $\omega^{\mu}$ which we denote as $X_{\mu}$. Thus $A_{v,\mu}=X_{\mu}A_{v}$. The electromagnetic portion of $\mathcal{S}_{matter}$ in (\ref{3})  is 
\begin{equation}\label{14}
\mathcal{S}_{matter}=\int dt dx^{3}N\sqrt{h}\left(-\frac{1}{16 \pi} F_{\mu \nu} F^{\mu \nu}\right),
\end{equation}
where 
\begin{equation}\label{15}
\begin{aligned}
h_{ab}=e^{2\alpha(t)+2\beta_{+}(t)}\operatorname{diag}\left(e^{2\sqrt{3}\beta_{-}(t)}, e^{-2\sqrt{3}\beta_{-}(t)}, e^{-6 \beta_{+}\left(t\right)}\right).
\end{aligned}
\end{equation}
Writing the action (\ref{14}) in terms of its vector potential $A$ and our structure constants results in the Lagrangian density which is derived in \cite{waller1984bianchi}
\begin{equation}\label{16}
\begin{aligned}
&\mathscr{L}=\Pi^{s} A_{0, s}-NH \\&
\mathscr{L}=\Pi^{s} A_{0, s}-\Pi^{s} A_{s, 0}-N\frac{2 \pi}{\sqrt{h}} \Pi^{s} \Pi^{p} h_{s p} \\& - \frac{N\sqrt{h}}{16 \pi} h^{i k} h^{s l}\left(2A_{[i,s]}+A_{m} C^{m}_{i s}\right)\left(2A_{[k,l]}+A_{m} C^{m}_{k l}\right),
\end{aligned}
\end{equation}
where
\begin{equation}\label{17}
\Pi^{s}=\frac{\partial \mathscr{L} }{\partial\left(X_{0} A_{s}\right)}=\frac{ h^{s j}\sqrt{h}}{4 N\pi} \left(-A_{0, j}+A_{j, 0}+A_{\alpha} C^{\alpha}_{0 j}\right),
\end{equation}
and we allow the shift $N^{k}$ to vanish. If we invoke the homogeneity of (\ref{1}) then we can say that $A_{i, j}=0$, and $ A_{0, j}=0$ which results in (\ref{16}) simplifying to 

\begin{equation}\label{18}
\begin{aligned}
&\mathscr{L}=\Pi^{s} A_{s, 0} \\&-N\left[\frac{2 \pi}{\sqrt{h}} \Pi^{s} \Pi^{p} h_{p s}+\frac{\sqrt{h}}{16 \pi} h^{i k} h^{s l} C^{m}_{k l}C^{n}_{i s} A_{m} A_{n}\right].
\end{aligned}
\end{equation}
The non zero Bianchi VIII and  IX structure constants are the following respectively

\begin{equation}\label{19}
\begin{aligned}
&C^{1}_{2 3}=-1 \\&
C^{1}_{3 2}=1 \\&
C^{2}_{1 3}=-1 \\&
C^{2}_{3 1}=1 \\&
C^{3}_{1 2}=1 \\&
C^{3}_{2 1}=-1.
\end{aligned}
\end{equation}

\begin{equation}\label{20}
\begin{aligned}
C_{j k}^{i}=\epsilon_{ijk}
\end{aligned}
\end{equation}

We will now set $A_{2}$, $A_{3}$, $\Pi^{2}$, and $\Pi^{3}$ to zero  and only consider the electromagnetic field produced by $A_{1}$ and $\Pi^{1}$; doing so results in the following Lagrangian density
\begin{equation}\label{21}
\begin{aligned}
&\mathscr{L}=\Pi^{1} A_{1,0} \\&-N\left[\frac{2 \pi}{\sqrt{h}} \Pi^{1} \Pi^{1} h_{1 1}+\frac{\sqrt{h}}{16 \pi} h^{i k} h^{s l} C^{1}_{k l}C^{1}_{i s} A_{1} A_{1}\right].
\end{aligned}
\end{equation}
As the reader can easily verify $ h^{i k} h^{s l} C^{1}_{k l}C^{1}_{i s}=\frac{2h_{11}}{h}$, which allows us to obtain the following set of Maxwell's equations when $A_{1}$ and $\Pi^{1}$ are varied
\begin{equation}\label{22}
\dot{A}_{1}-4 \pi \frac{1}{\sqrt{h}} \Pi^{1}h_{11}=0
\end{equation}
and 
\begin{equation}\label{23}
\dot{\Pi}^{1}+\frac{1}{4 \pi} \frac{1}{\sqrt{h}} h_{1 1} A_{1}=0.
\end{equation}
For the last equation we applied an integration by parts to the term $\Pi^{1} A_{1, 0}$ and dropped the total derivative term which vanishes at the spatial boundary. The solutions for (\ref{22}) and (\ref{23}) are 
\begin{equation}\label{24}
A_{1}=\sqrt{2}B_{0}cos(\theta(t))
\end{equation}
\begin{equation}\label{25}
\Pi^{1}=\frac{1}{2\sqrt{2}\pi}B_{0}sin(\theta(t))
\end{equation}
where $\theta(t)$ is an integral which is immaterial for our purposes and $B_{0}$ is an integration constant. Inserting (\ref{24}) and (\ref{25}) back into (\ref{21}) results in 
\begin{equation}\label{26}
\begin{aligned}
\mathscr{L}=\Pi^{1} &A_{1,0}-N\frac{B_{0}^{2}}{4\pi\sqrt{h}}h_{11}\left(sin(\theta(t))^{2}+cos(\theta(t))^{2}\right)\\&  =\Pi^{1} A_{1,0}-N\frac{B_{0}^{2}}{4\pi}e^{-\alpha(t)+2\beta(t)_{+}+2\sqrt{3}\beta(t)_{-}},
\end{aligned}
\end{equation}
From (\ref{16}) we can easily identify the electromagnetic Hamiltonian as 
\begin{equation}\label{27}
\begin{aligned}
H_{em}=\frac{B_{0}^{2}}{4\pi}e^{-\alpha(t)+2\beta(t)_{+}+2\sqrt{3}\beta(t)_{-}}
\end{aligned}
\end{equation}
which can be added to the Hamiltonian constraint which is derivable from our action (\ref{3})
\begin{equation}\label{28}
\begin{aligned}
&e^{-3 \alpha(t)}\left(-p_{\alpha}^{2}+p_{+}^{2}+p_{-}^{2}+12p_{\phi}^{2}\right)+U_{g} \\&+\frac{B_{0}^{2}}{4\pi}e^{-\alpha(t)+2\beta(t)_{+}+2\sqrt{3}\beta(t)_{-}}=0,
\end{aligned}
\end{equation}
where $U_{g}$ is the gravitational component of the potential. 
Quantizing (\ref{28}) using the Hartle-Hawking operator ordering, multiplying each side by $e^{3\alpha(t)}$, and rescaling $B_{0}$ results in one of the the Wheeler DeWitt equations of (\ref{7}). 

As the reader can verify for the cases when $A_{1}=0$, $A_{3}=0$, $\Pi^{1}=0$, and $\Pi^{3}=0$ the resulting electromagnetic term is 
\begin{equation}\label{29}
\begin{aligned}
U_{2 \hspace{1mm} em }=2b^{2}e^{2\alpha+2\beta_{+}-2\sqrt{3}\beta_{-}}.
\end{aligned}
\end{equation}
and the case when $A_{1}=0$, $A_{2}=0$, $\Pi^{1}=0$, and $\Pi^{2}=0$ results in the term reported in \cite{waller1984bianchi}
\begin{equation}\label{30}
\begin{aligned}
U_{3 \hspace{1mm} em }=2b^{2}e^{2\alpha-4\beta_{+}}.
\end{aligned}
\end{equation}
For the Taub models there are only two unique electromagnetic field potentials terms, (\ref{30}) and (\ref{29}) when $\beta_{-}=0.$

If we start with (\ref{21}) and directly quantize our component of the total Hamiltonian constraint which is proportional to the lapse N we obtain a similar, but slightly different contribution to the potential. Simplifying the term in brackets of (\ref{21}) results in the following contribution to the Hamiltonian constraint which can be derived with (\ref{3})

\begin{equation}\label{31}
\begin{aligned}
H_{em}=N\left[\frac{e^{-\alpha+2\beta_{+}+2\sqrt{3}\beta_{-}}\left(16 \pi ^2 \Pi^{1}\Pi^{1}+A^{2}_{1}\right)}{8 \pi }\right].
\end{aligned}
\end{equation}
The term $\left(16 \pi ^2 \Pi^{1}\Pi^{1}+A^{2}_{1}\right)$ commutes with our total Hamiltonian constraint $H_{gravity} +H_{em}+H_{scalar}$. Thus we can solve the following WDW equation constructed by directly quantizing (\ref{31}) with the rest of our constraint 

\begin{equation}\label{32}
\begin{aligned}
&\square \Psi-B \frac{\partial \Psi}{\partial \alpha}-e^{2\alpha+2\beta_{+}+2\sqrt{3}\beta_{-}}\left(-2\pi \frac{\partial^2 \Psi}{\partial A^{2}_{1}}+\frac{1}{8\pi}A^{2}_{1}\Psi\right)\\&+U_{\pm} \Psi=0,
\end{aligned}
\end{equation}
by solving this eigenvalue problem 
\begin{equation}\label{33}
\begin{aligned}
-2\pi \frac{\partial^2 \Psi}{\partial A^{2}_{1}}+\frac{1}{8\pi}A^{2}_{1}\Psi=b_{n}\Psi.
\end{aligned}
\end{equation}
This is simply the Schr$\text{\" o}$dinger equation for a harmonic oscillator whose solutions are well known
\begin{equation}\label{34}
\begin{aligned}
&\Psi=\psi\left(\alpha,\beta_{+},\beta_{-},\phi\right)e^{-\frac{A^{2}_{1}}{8 \pi }} H_{b_{n}-\frac{1}{2}}\left(\frac{A_{1}}{2 \sqrt{\pi }}\right)\\& b_{n}=\frac{1}{2}\left(1+2n\right).
\end{aligned}
\end{equation}
Inserting our $\Psi$ from (\ref{34}) into (\ref{32}) yields 
\begin{equation}\label{35}
\square \psi-B \frac{\partial \psi}{\partial \alpha}-b_{n}e^{2\alpha+2\beta_{+}+2\sqrt{3}\beta_{-}}+U_{\pm} \psi=0.
\end{equation}
This WDW equation is similar to what we had before except for the fact that the strength $b_{n}$ of the electromagnetic field is now quantized thanks to (\ref{32}). By first solving the classical $A_{i}$ equations (\ref{22}) in terms of the Misner variables we eliminate the electromagnetic field degree of freedom. For now though working with just (\ref{28}), (\ref{29}), and (\ref{30}) is sufficient for what will follow.

\section{\label{sec:level4}The Euclidean-signature semi classical method} 
Our outline of this method will follow closely \cite{moncrief2014euclidean}. The method described in this section and its resultant equations can in principle be used to find solutions and quantum corrections to a wide class of quantum cosmological models such as all of the Bianchi A, Kantowski Sachs models, and the FLRW models.  
 
The first step we will take in analyzing our WDW equations is to introduce the ansatz
 \begin{equation}\label{36}
\stackrel{(0)}{\Psi}_{\hbar}=e^{-S_{\hbar} / \hbar}
\end{equation}
where $S_{\hbar}$ is a function of $\left(\alpha,\beta_+,\beta_-,
\phi\right)$. We will rescale $S_{\hbar}$ in the following way  
\begin{equation}\label{37}
\mathcal{S}_{\hbar} :=\frac{G}{c^{3} L^{2}} S_{\hbar}
\end{equation}
where $\mathcal{S}_{\hbar}$ is dimensionless and admits the following power series in terms of this dimensionless parameter
\begin{equation}\label{38}
X :=\frac{L_{\text { Planck }}^{2}}{L^{2}}=\frac{G \hbar}{c^{3} L^{2}}.
\end{equation}
The series is given by 
\begin{equation}\label{39}
\mathcal{S}_{\hbar}=\mathcal{S}_{(0)}+X \mathcal{S}_{(1)}+\frac{X^{2}}{2 !} \mathcal{S}_{(2)}+\cdots+\frac{X^{k}}{k !} \mathcal{S}_{(k)}+\cdots,
\end{equation}
and as a result our initial ansatz now takes the following form 
\begin{equation}\label{40}
\stackrel{(0)}{\Psi}_{\hbar}=e^{-\frac{1}{X} \mathcal{S}_{(0)}-\mathcal{S}_{(1)}-\frac{X}{2 !} \mathcal{S}_{(2)}-\cdots}.
\end{equation}
Substituting this ansatz into the Wheeler-DeWitt equation and
requiring satisfaction, order-by-order in powers of X leads immediately to the sequence of equations

\begin{equation}\label{41}
\begin{aligned}
&\left(\frac{\partial \mathcal{S}_{(0)}}{\partial \alpha}\right)^{2}-\left(\frac{\partial \mathcal{S}_{(0)}}{\partial \beta_{+}}\right)^{2}-\left(\frac{\partial \mathcal{S}_{(0)}}{\partial \beta_{-}}\right)^{2}-12\left(\frac{\partial \mathcal{S}_{(0)}}{\partial \phi}\right)^{2} \\&+U_{\pm}=0
\end{aligned}
\end{equation}
\begin{equation}\label{42}
\begin{aligned}
& 2\Biggl[\frac{\partial \mathcal{S}_{(0)}}{\partial \alpha} \frac{\partial \mathcal{S}_{(1)}}{\partial \alpha}-\frac{\partial \mathcal{S}_{(0)}}{\partial \beta_{+}} \frac{\partial \mathcal{S}_{(1)}}{\partial \beta_{+}}-\frac{\partial \mathcal{S}_{(0)}}{\partial \beta_{-}} \frac{\partial \mathcal{S}_{(1)}}{\partial \beta_{-}} \\& -12\frac{\partial \mathcal{S}_{(0)}}{\partial \phi} \frac{\partial \mathcal{S}_{(1)}}{\partial \phi}\Biggr]  +B \frac{\partial \mathcal{S}_{(0)}}{\partial \alpha}-\frac{\partial^{2} \mathcal{S}_{(0)}}{\partial \alpha^{2}} \\&+\frac{\partial^{2} \mathcal{S}_{(0)}}{\partial \beta_{+}^{2}}+\frac{\partial^{2} \mathcal{S}_{(0)}}{\partial \beta_{-}^{2}}+12\frac{\partial^{2} \mathcal{S}_{(0)}}{\partial \phi^{2}}=0,
\end{aligned}
\end{equation},
\begin{equation}\label{43}
\begin{aligned}
& 2\Biggl[\frac{\partial \mathcal{S}_{(0)}}{\partial \alpha} \frac{\partial \mathcal{S}_{(1)}}{\partial \alpha}-\frac{\partial \mathcal{S}_{(0)}}{\partial \beta_{+}} \frac{\partial \mathcal{S}_{(1)}}{\partial \beta_{+}}-\frac{\partial \mathcal{S}_{(0)}}{\partial \beta_{-}} \frac{\partial \mathcal{S}_{(1)}}{\partial \beta_{-}} \\& -12\frac{\partial \mathcal{S}_{(0)}}{\partial \phi} \frac{\partial \mathcal{S}_{(1)}}{\partial \phi}\Biggr] +k\Biggl[B \frac{\partial \mathcal{S}_{(k-1)}}{\partial \alpha}-\frac{\partial^{2} \mathcal{S}_{(k-1)}}{\partial \alpha^{2}}\\&+ \frac{\partial^{2} \mathcal{S}_{(k-1)}}{\partial \beta_{+}^{2}}+\frac{\partial^{2} \mathcal{S}_{(k-1)}}{\partial \beta_{-}^{2}}+12\frac{\partial^{2} \mathcal{S}_{(k-1)}}{\partial \phi^{2}}\Biggr]  \\ & + \sum_{\ell=1}^{k-1} \frac{k !}{\ell !(k-\ell) !}\Biggr(\frac{\partial \mathcal{S}_{(\ell)}}{\partial \alpha} \frac{\partial \mathcal{S}_{(k-\ell)}}{\partial \alpha}-\frac{\partial \mathcal{S}_{(\ell)}}{\partial \beta_{+}} \frac{\partial \mathcal{S}_{(k-\ell)}}{\partial \beta_{+}}  \\&- \frac{\partial \mathcal{S}_{(\ell)}}{\partial \beta_{-}} \frac{\partial \mathcal{S}_{(k-\ell)}}{\partial \beta_{-}}- 12\frac{\partial \mathcal{S}_{(\ell)}}{\partial \phi} \frac{\partial \mathcal{S}_{(k-\ell)}}{\partial \phi}\Biggl) =0
\end{aligned}
\end{equation}

We will refer to $\mathcal{S}_{(0)}$ in our WDW wave functions as the leading order term, which can be used to construct approximate solutions to the Lorentzian signature WDW equation, and call $\mathcal{S}_{(1)}$ the first order term. The $\mathcal{S}_{(1)}$ term can also be viewed as our first quantum correction, with the other $\mathcal{S}_{(k)}$ terms being the additional higher order quantum corrections, assuming that they are smooth and globally defined. This is reflected in the fact that the higher order transport equations depend on the operator ordering used in defining the Wheeler Dewitt equation, which is an artifact of quantization. Additionally in some cases one can find a solution to the $\mathcal{S}_{(1)}$ equation which allows the $\mathcal{S}_{(2)}$ equation to be satisfied by zero. Then one can write down the following as a solution to the WDW equation for either a particular value of the Hartle-Hawking ordering parameter, or for an arbitrary ordering parameter depending on the $\mathcal{S}_{(1)}$ which is found.   

\begin{equation}\label{44}
\stackrel{(0)}{\Psi}_{\hbar}=e^{-\frac{1}{X} \mathcal{S}_{(0)}-\mathcal{S}_{(1)}}
\end{equation}.

This can be easily shown. Let's take $\mathcal{S}_{(0)}$ and $\mathcal{S}_{(1)}$ as arbitrary known functions which allow the $\mathcal{S}_{(2)}$ transport equation to be satisfied by zero, then the $k=3$ transport equation can be expressed as 
\begin{equation}\label{45}
\begin{aligned}
&2\Biggl[\frac{\partial \mathcal{S}_{(0)}}{\partial \alpha} \frac{\partial \mathcal{S}_{(3)}}{\partial \alpha}-\frac{\partial \mathcal{S}_{(0)}}{\partial \beta_{+}} \frac{\partial \mathcal{S}_{(3)}}{\partial \beta_{+}} \\&-\frac{\partial \mathcal{S}_{(0)}}{\partial \beta_{-}} \frac{\partial \mathcal{S}_{(3)}}{\partial \beta_{-}}-12\frac{\partial \mathcal{S}_{(0)}}{\partial \phi} \frac{\partial \mathcal{S}_{(3)}}{\partial \phi}\Biggr] =0
\end{aligned}
\end{equation}
which is clearly satisfied by $\mathcal{S}_{(3)}$=0. The $\mathcal{S}_{(4)}$ equation can be written in the same form as (\ref{45}) and one of its solution is 0 as well, thus resulting in the $\mathcal{S}_{(5)}$ equation possessing the same form as (\ref{45}). One can easily convince oneself that this pattern continues for all of the $k\geq 3$ $\mathcal{S}_{(k)}$ transport equations as long as the solution of the $\mathcal{S}_{(k-1)}$ transport equation is chosen to be 0. Thus if an $\mathcal{S}_{(1)}$ exists which allows one to set the solutions to all of the higher order transport equations to zero the infinite sequence of transport equations generated by our ansatz truncates to a finite sequence of equations which allows us to construct a closed form wave function satisfying the WDW equation. Not all solutions to the $\mathcal{S}_{(1)}$ transport equation will allow the $\mathcal{S}_{(2)}$ transport equation to be satisfied by zero; however in our case, we were able to find $\mathcal{S}_{(1)}$'s which cause the $\mathcal{S}_{(2)}$ transport equation to be satisfied by zero, thus allowing one to set all of the solutions to the higher order transport equations to zero as shown above. This will enable us to construct  closed form solutions to the Lorentzian signature Bianchi VIII and IX WDW equations. It should be noted that using an alternate form of operator ordering for the WDW equation which we will introduce later one can construct solutions to it using just the $\mathcal{S}_{(0)}$ term.

Under certain conditions which we cannot rigorously articulate yet, wave functions that behave as 'excited' states can be calculated by introducing the following ansatz. 
\begin{equation}\label{46}
{\Psi}_{\hbar}={\Phi}_{\hbar} e^{-S_{\hbar} / \hbar}
\end{equation}
where $$
S_{\hbar}=\frac{c^{3} L^{2}}{G} \mathcal{S}_{\hbar}=\frac{c^{3} L^{2}}{G}\left(\mathcal{S}_{(0)}+X \mathcal{S}_{(1)}+\frac{X^{2}}{2 !} \mathcal{S}_{(2)}+\cdots\right)
$$
is the same series expansion as before and ${\Phi}_{\hbar}$ can be expressed as the following series 
\begin{equation}\label{47}
{\Phi_{\hbar}=\Phi_{(0)}+X \Phi_{(1)}+\frac{X^{2}}{2 !} \Phi_{(2)}+\cdots+\frac{X^{k(*)}}{k !} \Phi_{(k)}+\cdots}
\end{equation}
with X being the same dimensionless quantity as before. 
Inserting (\ref{46}) with the expansions given by (\ref{39}) and (\ref{47}) into the Wheeler DeWitt equation (\ref{7}) and by matching equations in powers of X leads to the following sequence of equations. 
\begin{equation}\label{48}
\begin{aligned}
&-\frac{\partial \Phi_{(0)}}{\partial \alpha} \frac{\partial \mathcal{S}_{(0)}}{\partial \alpha}+\frac{\partial \Phi_{(0)}}{\partial \beta_{+}} \frac{\partial \mathcal{S}_{(0)}}{\partial \beta_{+}}\\&+\frac{\partial \Phi_{(0)}}{\partial \beta_{-}} \frac{\partial \mathcal{S}_{(0)}}{\partial \beta_{-}}+12\frac{\partial \Phi_{(0)}}{\partial \phi} \frac{\partial \mathcal{S}_{(0)}}{\partial \phi}=0,
\end{aligned}
\end{equation},
\begin{equation}\label{49}
\begin{aligned}
&-\frac{\partial \Phi_{(1)}}{\partial \alpha} \frac{\partial \mathcal{S}_{(0)}}{\partial \alpha}+\frac{\partial \Phi_{(1)}}{\partial \beta_{+}} \frac{\partial \mathcal{S}_{(0)}}{\partial \beta_{+}}\\&+\frac{\partial \Phi_{(1)}}{\partial \beta_{-}} \frac{\partial \mathcal{S}_{(0)}}{\partial \beta_{-}}+12\frac{\partial \Phi_{(1)}}{\partial \phi} \frac{\partial \mathcal{S}_{(0)}}{\partial \phi} \\ & +\Biggl(-\frac{\partial \Phi_{(0)}}{\partial \alpha} \frac{\partial \mathcal{S}_{(1)}}{\partial \alpha}+\frac{\partial \Phi_{(0)}}{\partial \beta_{+}} \frac{\partial \mathcal{S}_{(1)}}{\partial \beta_{+}}\\&+\frac{\partial \Phi_{(0)}}{\partial \beta_{-}} \frac{\partial \mathcal{S}_{(1)}}{\partial \beta_{-}}+12\frac{\partial \Phi_{(0)}}{\partial \phi} \frac{\partial \mathcal{S}_{(1)}}{\partial \phi}\Biggr) \\ & +\frac{1}{2}\Biggl(-B \frac{\partial \Phi_{(0)}}{\partial \alpha}+\frac{\partial^{2} \Phi_{(0)}}{\partial \alpha^{2}}-\frac{\partial^{2} \Phi_{(0)}}{\partial \beta_{+}^{2}} \\&-\frac{\partial^{2} \Phi_{(0)}}{\partial \beta_{-}^{2}}-12\frac{\partial^{2} \Phi_{(0)}}{\partial \phi^{2}}\Biggr)=0,
\end{aligned}
\end{equation}
\begin{equation}\label{50}
\begin{aligned}
& -\frac{\partial \Phi_{(k)}}{\partial \alpha} \frac{\partial \mathcal{S}_{(0)}}{\partial \alpha}+\frac{\partial \Phi_{(k)}}{\partial \beta_{+}} \frac{\partial \mathcal{S}_{(0)}}{\partial \beta_{+}}\\&+\frac{\partial \Phi_{(k)}}{\partial \beta_{-}} \frac{\partial \mathcal{S}_{(0)}}{\partial \beta_{-}}+12\frac{\partial \Phi_{(k)}}{\partial \phi} \frac{\partial \mathcal{S}_{(0)}}{\partial \phi} \\ & 
+k\Biggr(-\frac{\partial \Phi_{(k-1)}}{\partial \alpha} \frac{\partial \mathcal{S}_{(1)}}{\partial \alpha}+\frac{\partial \mathcal{S}_{(1)}}{\partial \beta_{+}} \frac{\partial \mathcal{S}_{(1)}}{\partial \beta_{+}} \\&+\frac{\partial \Phi_{(k-1)}}{\partial \beta_{-}} \frac{\partial \mathcal{S}_{(1)}}{\partial \beta_{-}}+12\frac{\partial \Phi_{(k-1)}}{\partial \phi} \frac{\partial \mathcal{S}_{(1)}}{\partial \phi}\Biggr) \\ & 
+\frac{k}{2}\Biggr(-B \frac{\partial \Phi_{(k-1)}}{\partial \alpha}+\frac{\partial^{2} \Phi_{(k-1)}}{\partial \alpha^{2}}-\frac{\partial^{2} \Phi_{(k-1)}}{\partial \beta_{+}^{2}}\\&-\frac{\partial^{2} \Phi_{(k-1)}}{\partial \beta_{-}^{2}}-12\frac{\partial^{2} \Phi_{(k-1)}}{\partial \phi^{2}}\Biggr)\\ & -
\sum_{\ell=2}^{k} \frac{k !}{\ell !(k-\ell) !}\Biggr( \frac{\partial \Phi_{(k-\ell)}}{\partial \alpha} \frac{\partial \mathcal{S}_{(\ell)}}{\partial \alpha}-\frac{\partial \Phi_{(k-\ell)}}{\partial \beta_{+}} \frac{\partial \mathcal{S}_{(\ell)}}{\partial \beta_{+}} \\&-  \frac{\partial \Phi_{(k-\ell)}}{\partial \beta_{-}} \frac{\partial \mathcal{S}_{(\ell)}}{\partial \beta_{-}}- 12 \frac{\partial \Phi_{(k-\ell)}}{\partial \phi} \frac{\partial \mathcal{S}_{(\ell)}}{\partial \phi}\Biggr) =0.
\end{aligned}
\end{equation}

It can be seen from computing $\frac{d\Phi_{(0)}\left(\alpha,\beta_+,\beta_-,\phi \right)}{dt}=\dot{\alpha}\frac{\partial \Phi_{(0)}}{\partial \alpha}+\dot{\beta_+}\frac{\partial \Phi_{(0)}}{\partial \beta_+}+\dot{\beta_-}\frac{\partial \Phi_{(0)}}{\partial \beta_-}+\dot{\phi}\frac{\partial \Phi_{(0)}}{\partial \phi}$, and extrapolating from $\left(4.9, \hspace{1 mm} 4.18-4.20\right)$ of  \cite{moncrief2014euclidean} what $\dot{\phi}$ is in terms of $\mathcal{S}_{(0)}$ that $\Phi_{(0)}$ is a conserved quantity under the flow of $S_{0}$. This means that any function $F\left(\Phi_{(0)}\right)$ is also a solution of equation (\ref{48}). Wave functions constructed from these functions of $\Phi_{0}$ are only physical if they are smooth and globally defined.  If we choose our $\phi_{0}$ to have the form $f\left(\alpha,\beta_+,\beta_-,\phi \right)^{m}$ where $f\left(\alpha,\beta_+,\beta_-,\phi \right)$ is some function which is conserved under the flow of $\mathcal{S}_{(0)}$ and vanishes for some finite values of the Misner variables then we must restrict $m$ to be either zero or a positive integer. For bound states $m$ can plausibly be interpreted as graviton excitation number \cite{bae2014quantizing}. This makes our 'excited' states akin to bound states in quantum mechanics like the harmonic oscillator whose excited states are denoted by discrete integers as opposed to a continuous index. This discretization of the quantity that denotes our 'excited' states is the mathematical manifestation of quantization one would expect excited states to possess in quantum dynamics.

 If our conserved quantity  $f\left(\alpha,\beta_+,\beta_-,\phi \right)$ does not vanish in minisuperspace then our 'excited' states can be interpreted as 'scattering' states akin to the quantum free particle and $m$ can be any real number. Beyond the leading order limit, if smooth globally defined solutions can be proven to exist for the higher order $\phi$ transport equations then one can construct asymptotic or closed form 'excited' state solutions for the quantum Taub models as we shall do. Additional information for why we can call the above 'excited' states despite them being solutions to an equation which does not have the same form as the Schr$\text{\" o}$dinger equation can be found in \cite{moncrief2014euclidean}. In what follows we will work in units in which $X=1$.

\section{\label{sec:level5}Bianchi VIII and IX Quantum Cosmology With An Aligned Electromagnetic Field, Scalar Field, Cosmological Constant, And, Stiff Matter} 
The author found the following solutions to the Euclidean-signature Hamilton Jacobi equations (\ref{41}) when an electromagnetic field, scalar field, cosmological constant, and stiff matter are present

\begin{equation}\label{51}
\begin{aligned}
&\mathcal{S}^{1}_{(0 \pm)}=\frac{1}{6} e^{2 \alpha-4\beta_{+}} \left(2 e^{6\beta_{+}} \cosh \left(2 \sqrt{3}
   \beta_{-}\right)\pm 1\right)\\&\mp\frac{\Lambda e^{4 \alpha+4 \beta_{+}}}{36 \pi }\mp\alpha \text{b}^2\mp\beta_{+}
   \text{b}^2+\frac{\phi \sqrt{\rho}}{2 \sqrt{3}},
\end{aligned}
\end{equation}

\begin{equation}\label{52}
\begin{aligned}
&\mathcal{S}^{2}_{(0 \pm)}=\frac{1}{6} e^{2 \alpha-4\beta_{+}} \left(2 e^{6\beta_{+}} \cosh \left(2 \sqrt{3}
   \beta_{-}\right)\pm 1\right)\\&-\frac{\Lambda e^{4 \alpha-2 \beta_{+}-2\sqrt{3}\beta_{-}}}{36 \pi }-\alpha \text{b}^2+\frac{\beta_{+}}{2}b^{2}+\frac{\sqrt{3}}{2}\beta_{-}
   \text{b}^2+\frac{\phi \sqrt{\rho}}{2 \sqrt{3}},
\end{aligned}
\end{equation}

\begin{equation}\label{53}
\begin{aligned}
&\mathcal{S}^{3}_{(0 \pm)}=\frac{1}{6} e^{2 \alpha-4\beta_{+}} \left(2 e^{6\beta_{+}} \cosh \left(2 \sqrt{3}
   \beta_{-}\right)\pm 1\right)\\&-\frac{\Lambda e^{4 \alpha+2 \beta_{+}-2\sqrt{3}\beta_{-}}}{36 \pi }-\alpha \text{b}^2+\frac{\beta_{+}}{2}b^{2}-\frac{\sqrt{3}}{2}\beta_{-}
   \text{b}^2+\frac{\phi \sqrt{\rho}}{2 \sqrt{3}}.
\end{aligned}
\end{equation}
In our solutions the plus $+$ sign or the top operation in $\pm$ and $\mp$ are for the Bianchi IX models while the bottom symbols/operators are for the Bianchi VIII models. It is interesting to note that in the limit of our matter sources vanishing these solutions approach the well known 'wormhole' \cite{moncrief1991amplitude} Bianchi IX solutions and its analogue  for the Bianchi VIII models. The author was unable to find elementary solutions to (\ref{41}) with the aforementioned matters sources that exhibited this property for the Bianchi IX 'no boundary'\cite{graham1993supersymmetric} or "arm" solutions \cite{barbero1996minisuperspace}, or for their Bianchi VIII analogues \cite{bene1994supersymmetric}. This potentially suggests that their is something special about the 'wormhole' solution. 

Using these expressions we can obtain a semi-classical solution to the WDW equation expressed in the Hartle-Hawking semi general operator ordering which respects the $\frac{2\pi}{3}$ symmetry in $\beta$ space present in the Bianchi IX potential when our electromagnetic field is zero($b=0$) or a solution to the WDW expressed in an alternative operator ordering when $b \ne 0$. 

As pointed out by Moncrief and Ryan in \cite{moncrief1991amplitude} and shown explicitly in \cite{giampieri1991new}, using a different operator ordering the for Wheeler DeWitt equation than the Hartle-Hawking one\cite{hartle1983wave} allows one to construct wave functions which satisfy it if one possesses pure imaginary solutions to its corresponding Lorentzian signature Hamilton Jacobi equation. We will review the derivation presented in \cite{giampieri1991new} which allows us to construct solutions using just an $\mathcal{S}_{(0)}$. The Hamiltonian constraint for the Bianchi A models can be expressed as 
\begin{equation}\label{54}
H=G^{A B} p_{A} p_{B}+U=0
\end{equation}
where $G^{A B}$ is the DeWitt supermetric and $p_{i}$ are the canonical momenta. Likewise the regular Lorentzian signature Hamiliton Jacobi is expressed as 
\begin{equation}\label{55}
G^{A B} \frac{\partial J}{\partial x^{A}} \frac{\partial J}{\partial x^{B}}+U=0.
\end{equation}
Because (\ref{55}) is the Lorentzian signature Hamilton Jacobi equation the signs in its derivatives are the opposite of those for the Euclidean case. That means for the Bianchi VIII and IX models with a cosmological constant, a primordial electromagnetic field, scalar field, and stiff matter (\ref{55}) is satisfied by our previous solutions (\ref{51}), (\ref{52}), and (\ref{53}) multiplied by $\sqrt{-1}$ such that $J=\sqrt{-1}\mathcal{S}^{i}_{( 0)}$. This allows us to rewrite (\ref{55})  as 
\begin{equation}\label{56}
\begin{aligned}
G^{A B} p_{A} p_{B}+G^{A B} \frac{\partial \mathcal{S}^{i}}{\partial x^{A}} \frac{\partial \mathcal{S}^{i}}{\partial x^{B}} 
=G^{A B}(x) \pi_{A}^{*} \pi_{B}=0
\end{aligned}
\end{equation}
where 
\begin{equation}\label{57}
\pi_{A}=p_{A}-\sqrt{-1} \frac{\partial  \mathcal{S}^{i}}{\partial x^{A}}
\end{equation} 
and is quantized as follows 
\begin{equation}\label{58}
\hat{\pi}_{A}=-\sqrt{-1} \frac{\partial}{\partial x^{A}}-\sqrt{-1} \frac{\partial  \mathcal{S}^{i}}{\partial x^{A}}.
\end{equation}
Due to this quantization, wave functions of the form $\Psi=e^{-\mathcal{S}^{i}_{(0)}}$ mathematically behave in the following way 
\begin{equation}\label{59}
\hat{\pi}_{B} \Psi=0.
\end{equation}
Thus we can satisfy the Bianchi VIII and IX WDW equations when a cosmological constant, a primordial electromagnetic field, scalar field, and stiff matter are present if we order the WDW as follows 

\begin{equation}\label{60}
\frac{1}{\sqrt{|G|}}\left[\hat{\pi}_{A}^{*}\left(\sqrt{|G|} G^{A B} \hat{\pi}_{B}\right)\right]\Psi=0.
\end{equation}

We will first study a semi-classical solution \begin{equation}\label{61}
\Psi=\frac{1}{3}\left(e^{-\mathcal{S}^{1}_{(0 \hspace{1mm}  +)}}+e^{-\mathcal{S}^{2}_{(0 \hspace{1mm}  +)}}+e^{-\mathcal{S}^{3}_{(0 \hspace{1mm} +)}}\right)
\end{equation}
to the WDW equation(\ref{7}) when $b =0$ and then turn our attention to a closed form solution which satisfies (\ref{60}) when $b \ne0$. As we previously mentioned there are two candidates for our evolution parameter $\alpha$ and $\phi$. Even though $\phi$ is in some respects a better variable because classically it is guaranteed to increase monotonically for the Bianchi VIII and IX models we are considering, we will analyze (\ref{61}) using $\alpha$ because it better facilitates the author's ability to convey the physical implications of our wave functions for this particular case. Below are three plots of (\ref{61}) for three different values of $\alpha$.

\onecolumngrid\
\begin{center}
\begin{figure}[h]
\centering
\begin{subfigure}{.4\textwidth}
  \centering
  \includegraphics[scale=.155 ]{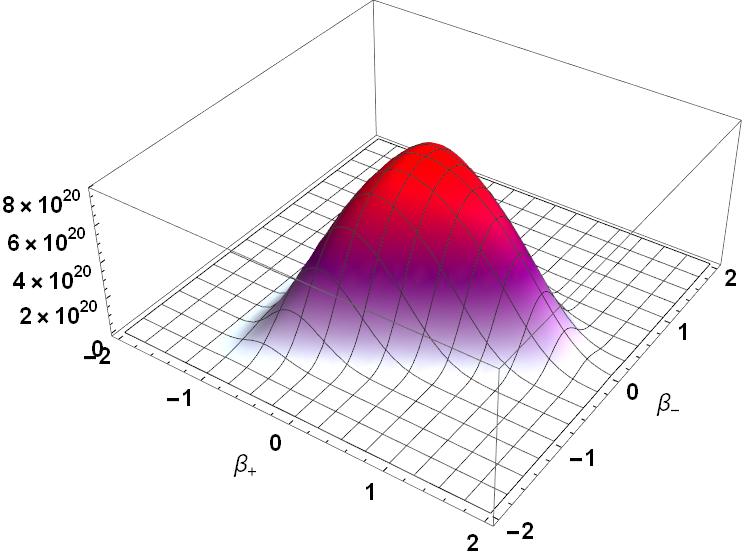}
  \caption{ $\alpha=-1$ \hspace{1mm} $\Lambda=-1$}
  \label{1a}
\end{subfigure}%
\begin{subfigure}{.4\textwidth}
  \centering
  \includegraphics[scale=.155 ]{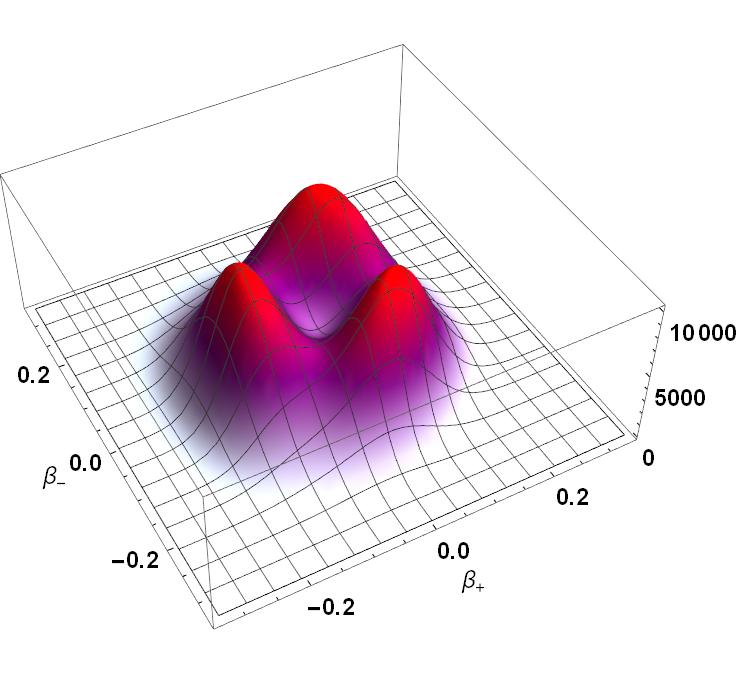}
  \caption{ $\alpha=1.65$ \hspace{1mm} $\Lambda=-1$}
  \label{1b}
\end{subfigure}
\begin{subfigure}{.4\textwidth}
  \centering
 \ \includegraphics[scale=.155]{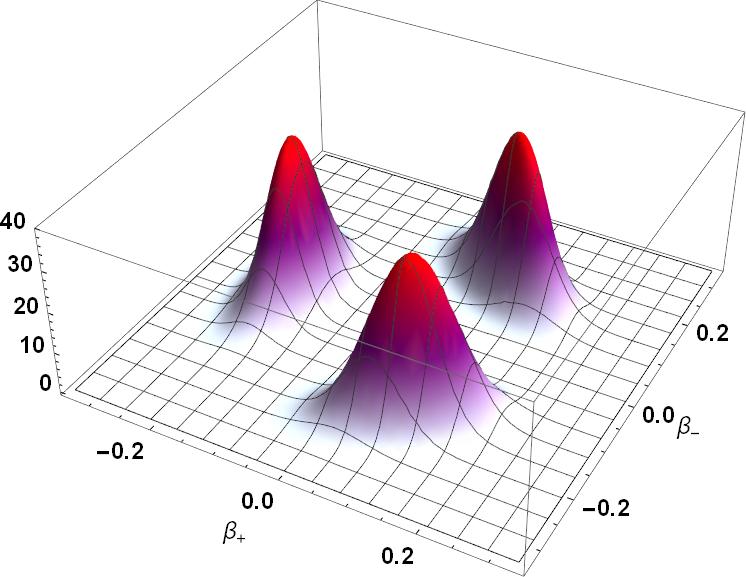}
  \caption{ $\alpha=2$ \hspace{1mm} $\Lambda=-1$}
  \label{1c}
\end{subfigure}%
\caption{Three different plots of (\ref{61}) for three different values of $\alpha$ where we suppress the the $\phi$ degree of freedom.}
\label{fig1}
\end{figure}
\end{center}
\twocolumngrid\

From here one out we will interpret the peaks which appear in our wave functions as geometric states that our quantum universes can tunnel in and out of as was done in \cite{garcia2002noncommutative}.
When $\alpha$ is less than 1.6 our wave function's peak is located at the origin in $\beta$ space which corresponds to isotropy. For negative values of $\alpha$ our wave function is roughly spread evenly around the origin as can be seen in figure (\ref{1a}). Thus we can say that figure (\ref{1a}) describes a universe with a "fuzzy" geometry. As $\alpha$ grows though our wave function becomes more sharply peaked at the origin. This is expected because geometrical "fuzziness" described by a wave function which is spread out in $\beta$ space is a quantum effect we intuitively expect to diminish as the universe grows in size dictated by $\alpha$. 

However around $\alpha \approx 1.6$ a dent forms where the wave function used to have an isotropic peak as can be seen in (\ref{1b}) and our wave function begins to split apart. This is a result of the influence of our negative cosmological constant $\Lambda$. As $\alpha$ continues to grow our wave function splits into three parts whose peaks are not centered around the origin. This indicates that the cosmological constant for our wave function acts as a driver of anisotropy by tearing our wave functions away from the origin in $\beta$ space. This is reminiscent of how a positive cosmological constant causes distant objects which are not gravitational bound to each other in the universe to accelerate away from each other. Of course this is just an analogy because our wave function is in minisuperspace, not space-time. 

The wave function shown in (\ref{1c}) is aesthetically similar to the wave functions\cite{paternoga1996exact} which were computed by employing Chern-Simons solutions in Ashtekar’s variables. Investigating the connection between the caustics which were studied in \cite{paternoga1996exact} and these elementary closed form solutions to the Euclidean-signature Hamilton Jacobi equation could potentially yield some interesting results. 

In comparison to the exotic and technical methods which were used to compute Bianchi IX wave functions with a negative cosmological constant in \cite{paternoga1996exact}, the fact that the Euclidean-signature semi classical method allowed us to compute similar Bianchi IX wave functions in closed form and expressed in terms of elementary functions is an impressive feat. This provides further support that the Euclidean-signature semi classical method is an effective way to prove the existence of solutions to Lorentzian signature equations. As is discussed in section 7 of \cite{moncrief2014euclidean}, it may be easier to prove the existence of formal solutions to the full functional Lorentzian signature Wheeler DeWitt equation using the Euclidean-signature Einstein-Hamilion-Jacobi(EHJ) equation than it is through the Lorentzian signature (EHJ). By successfully employing\cite{bae2015mixmaster,berkowitz2020towards,berkowitz2020atowards} this method in quantum cosmology we further support the idea that this method can in some circumstances be a useful alternative\cite{moncrief2014euclidean} to Euclidean path integrals\cite{gibbons1979positive,gibbons1993euclidean} for tackling problems in quantum gravity. More information on how the Euclidean-signature semi classical method applies to quantum cosmology, quantum gravity, and a plethora of field theories, including the Yang Mills 'mass gap' problem is provided here\cite{moncrief2012modified,moncrief2014euclidean, marini2016euclidean,marini2019euclidean,marini2020euclidean}.

\begin{center}
\begin{figure}[h]
\centering
\begin{subfigure}{.4\textwidth}
  \centering
  \includegraphics[scale=.2 ]{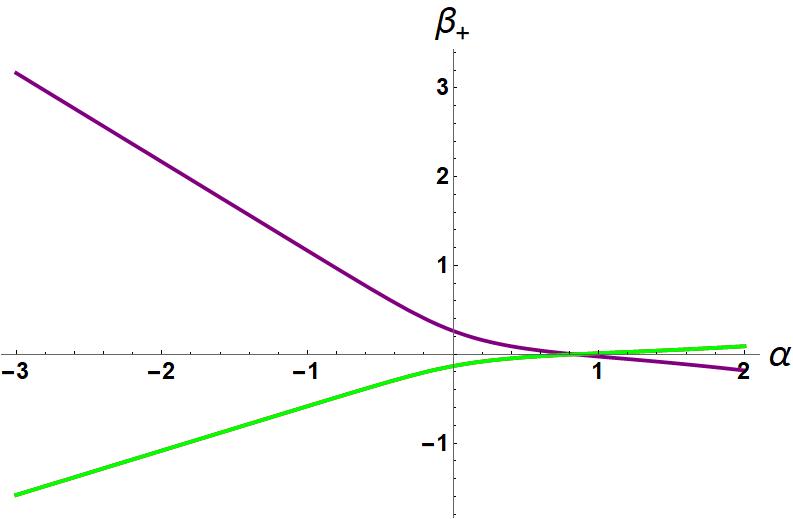}
  \caption{ $\Lambda=-1$}
  \label{2a}
\end{subfigure}%
\begin{subfigure}{.4\textwidth}
  \centering
  \includegraphics[scale=.2 ]{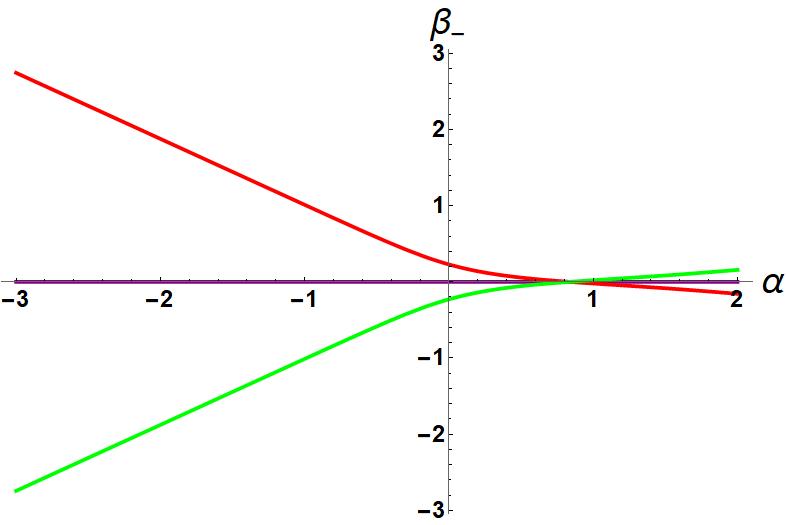}
  \caption{ $\Lambda=-1$}
  \label{2b}
\end{subfigure}
\caption{Two plots of the wave functions constructed from (\ref{51}), (\ref{52}), and (\ref{53}) which show their maximum values in $\beta$ space as a function of $\alpha$. The red line corresponds to (\ref{53}), the green line corresponds to (\ref{52}), and the purple line corresponds to (\ref{51}).}
\label{fig2}
\end{figure}
\end{center}

Another way of visualizing how the cosmological constant is a driver of anisotropy in our wave functions is through the plots in figure (\ref{fig2}).

\onecolumngrid\
\twocolumngrid\
Moving on to the case when an electromagnetic field is present we will analyze the solution to (\ref{60}) constructed from (\ref{51}). We don't lose much from not analyzing (\ref{52}) and (\ref{53}) because those are just (\ref{51}) rotated by $\pm \frac{2\pi}{3}$ in $\beta$ space. In figure 3 are four plots which showcase the possible effects that an aligned electromagnetic field can have on our particular 'ground' state wave functions.

\begin{center}
\begin{figure}[h]
\centering
\begin{subfigure}{.4\textwidth}
  \centering
  \includegraphics[scale=.156 ]{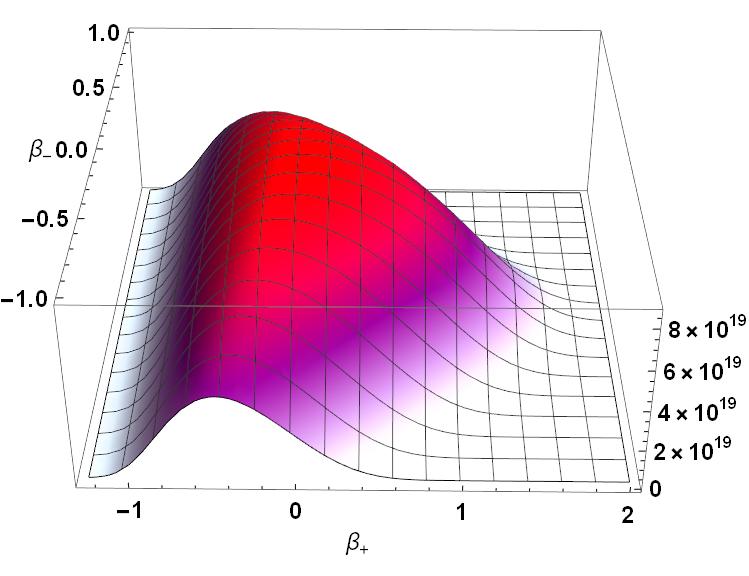}
  \caption{ $\alpha=-1$ \hspace{1mm} $\Lambda=-1$  \hspace{1mm} $b=0$}
  \label{3a}
\end{subfigure}%
\begin{subfigure}{.4\textwidth}
  \centering
  \includegraphics[scale=.156 ]{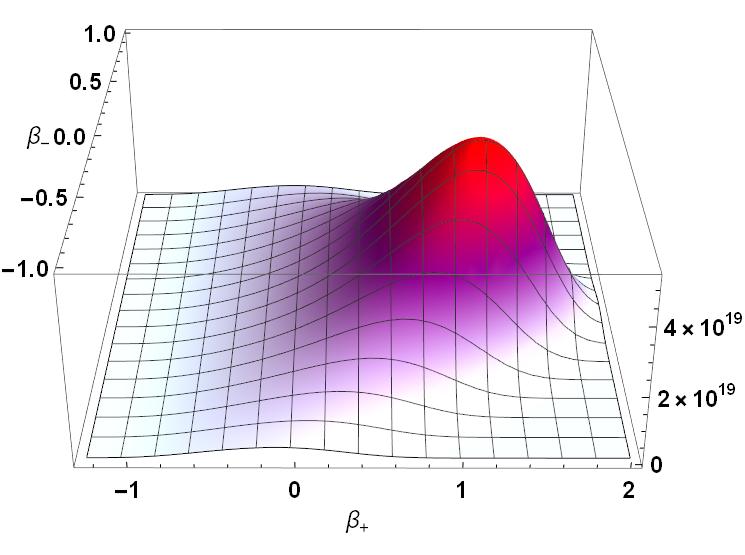}
  \caption{ $\alpha=-1$ \hspace{1mm} $\Lambda=-1$  \hspace{1mm} $b=1$}
  \label{3b}
\end{subfigure}
\begin{subfigure}{.4\textwidth}
  \centering
  \includegraphics[scale=.156]{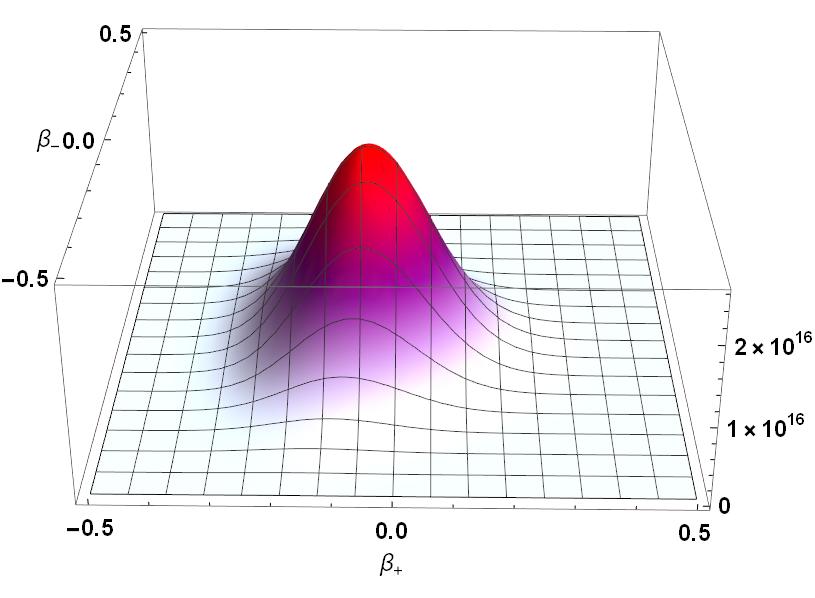}
  \caption{ $\alpha=1$ \hspace{1mm} $\Lambda=-1$  \hspace{1mm} $b=0$}
  \label{3c}
\end{subfigure}
\begin{subfigure}{.4\textwidth}
  \centering
 \ \includegraphics[scale=.126]{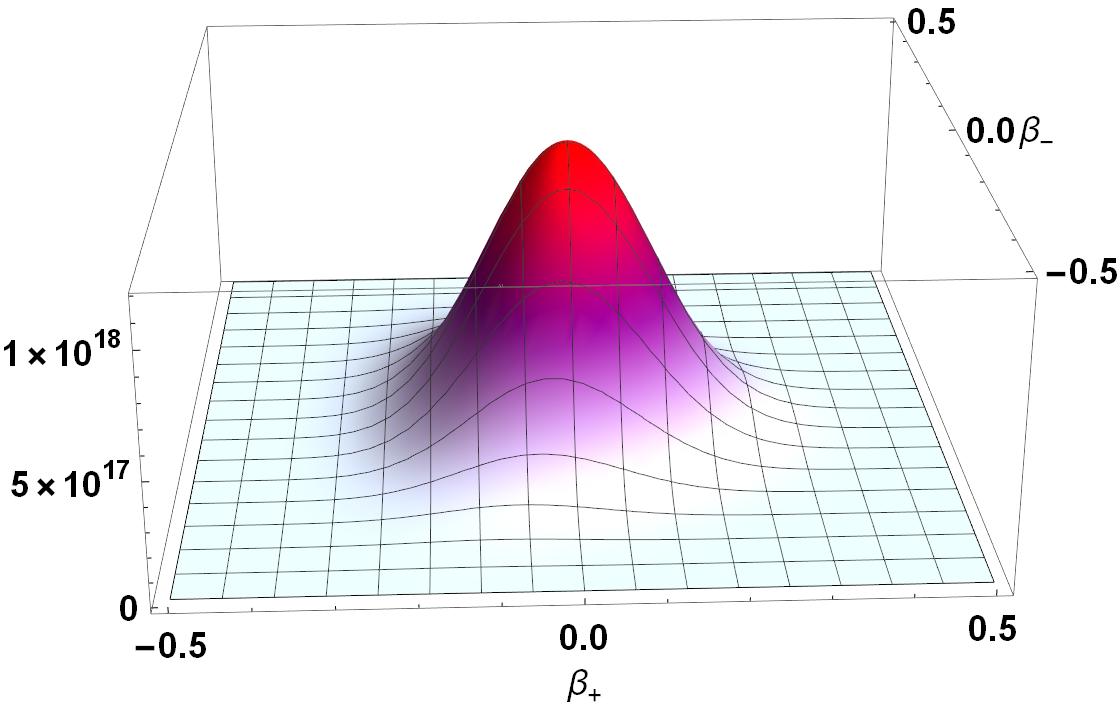}
  \caption{ $\alpha=1$ \hspace{1mm} $\Lambda=-1$  \hspace{1mm} $b=1.4$}
  \label{3d}
\end{subfigure}%
\caption{These are four plots of our wave functions constructed from (\ref{51}) with listed values for their cosmological constant and electromagnetic field.} 
\label{fig3}
\end{figure}
\end{center}

As can be seen by comparing figures (\ref{3a}) and (\ref{3b}) one of the effects of our electromagnetic field is that it causes our wave function to form a peak at a larger value of $\beta_{+}$ than it would otherwise. However owing to the primordial nature of our electromagnetic field for large values of $\alpha$ this effect is far less drastic as can be seen by comparing figures  (\ref{3c}) and (\ref{3d}). This suggest that an electromagnetic field can play a decisive role in increasing the prevalence of anisotropy in the early universe but still allow for a universe which becomes roughly isotropic as it continues to grow in size. To test this assertion, in the future we will need to study other quantum cosmological models with more general electromagnetic fields. In addition we would need to add inhomogeneities to those models. For now this finding contributes towards a theoretical understanding of how a primordial electric/magnetic field could have influenced the seeds of anisotropy in the early universe and how those seeds developed as it grew in size. 

Another noticeable effect of our electromagnetic field is that it causes our wave function to be more sharply peaked at smaller values of $\alpha$ than it otherwise would be as can be seen by comparing figures (\ref{3a}) and (\ref{3b}). This suggest that an electromagnetic field could play a role in causing a universe which initially possesses a "fuzzy" geometry in the sense that its wave function of the universe is evenly spread over $\beta$ space to transition to one which is sharply peaked at a particular point in $\beta$ space. This transition from a "fuzzy" geometry to a semi-classical one could occur if the primordial electromagnetic field was something that emerged in an early universe at some point in its evolution or was present in the beginning, but its effects were initially suppressed by Planck or GUT level physics. Such a mechanism being present in general anisotropic quantum cosmologies can help explain how a universe can transition from a state where it can only be accurately described using quantum mechanics to one which can be adequately described by classical mechanics.

\begin{center}
\begin{figure}[h]
\centering
\begin{subfigure}{.4\textwidth}
  \centering
  \includegraphics[scale=.15]{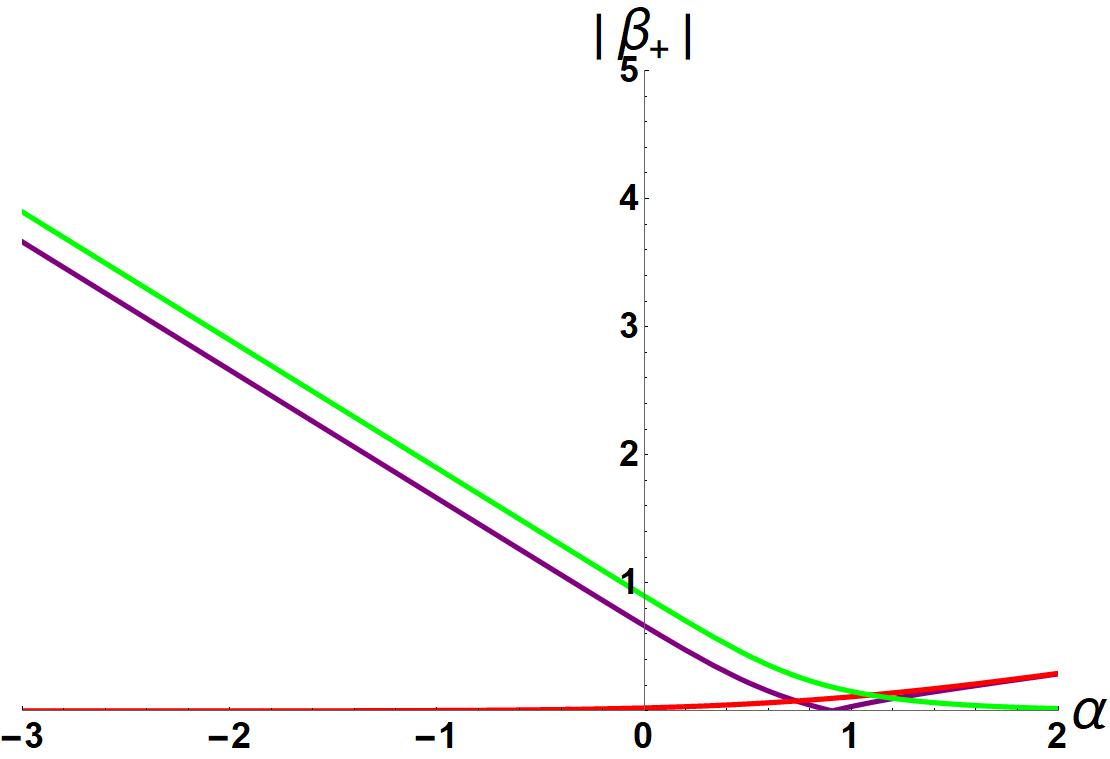}
  \caption{ $\alpha=-1$ \hspace{1mm} $\Lambda=-1$  \hspace{1mm} $b=0$}
  \label{4a}
\end{subfigure}%
\caption{This plot shows the maximum values of our wave functions in $\beta$ space as a function of $\alpha$. The green plot is of the wave function constructed from (\ref{51}) when no matter sources are present. The purple line is when both a negative cosmological constant $\Lambda=-3$ and an electromagnetic field $b=2$ are present. The red line is when just an electromagnetic field $b=2$ is present. }
\label{fig4}
\end{figure}
\end{center}

A surprising effect of our electromagnetic field is that it can actually reduce the level of anisotropy of a quantum universe at certain values of $\alpha$. This can be seen by comparing figures  (\ref{3c}) and (\ref{3d}). Our cosmological constant has a proclivity towards causing our wave functions to be centered at a negative value of $\beta_{+}$ while our aligned electromagnetic field causes our wave functions to shift toward the positive portion of the $\beta_{+}$ axis. Thus for certain values of $\alpha$ these two competing effects to increase anisotropy can cancel each other out and result in a reduction of anisotropy. This effect may also be caused by an aligned electromagnetic field and other forms of matter. Visually this dual nature concerning the tendencies of our two matter sources to induce anisotropy in our quantum cosmological can be visualized in figure (\ref{fig4}).

\onecolumngrid\

\twocolumngrid\

 We should note that this dual nature may be a result of us considering an aligned electromagnetic field and not one which has components in multiple directions. When both a negative cosmological constant and an electromagnetic field are present, isotropy is more likely to be reached at a smaller value of $\alpha$ than it is when either one or none of the matter sources are present. When an aligned electromagnetic field and a negative cosmological constant are present the magnitude of our wave functions decay as $\alpha$ continues to grow past a point. These results are in concord with those that the author observed in his previous works where he studied the effects of aligned electromagnetic fields on Taub\cite{berkowitz2020towards}, Bianchi II, and VII$h_{0}$\cite{berkowitz2020atowards} quantum cosmologies. 

\section{\label{sec:level6} IX Quantum Cosmology On The $\beta_{+}$ Axis}

In this section we will further explore Bianchi IX quantum cosmologies with matter sources by analyzing wave functions which correspond to classical trajectories in minisuperspace where $\beta_{-}$ starts at a fixed point in $\beta$ space. To start we need to pick an $\mathcal{S}_{(0)}$ which induces a flow in minisuperspace which possesses fixed points. To determine this we will write out the classical flow equation for an arbitrary $\mathcal{S}_{(0)}$
\onecolumngrid\
\begin{center}
\begin{figure}[h]
\centering
\begin{subfigure}{.4\textwidth}
  \centering
  \includegraphics[scale=.162]{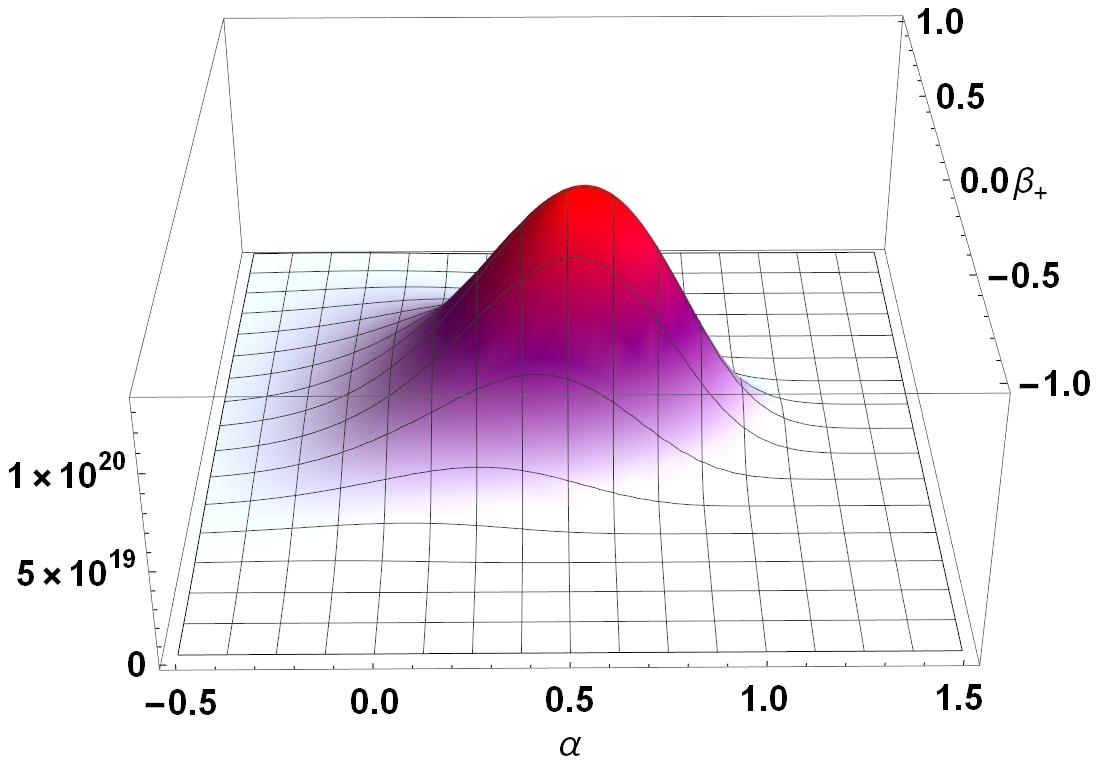}
  \caption{$\phi=0$ \hspace{1mm} $b=0$}
  \label{5a}
\end{subfigure}%
\begin{subfigure}{.4\textwidth}
  \centering
  \includegraphics[scale=.144 ]{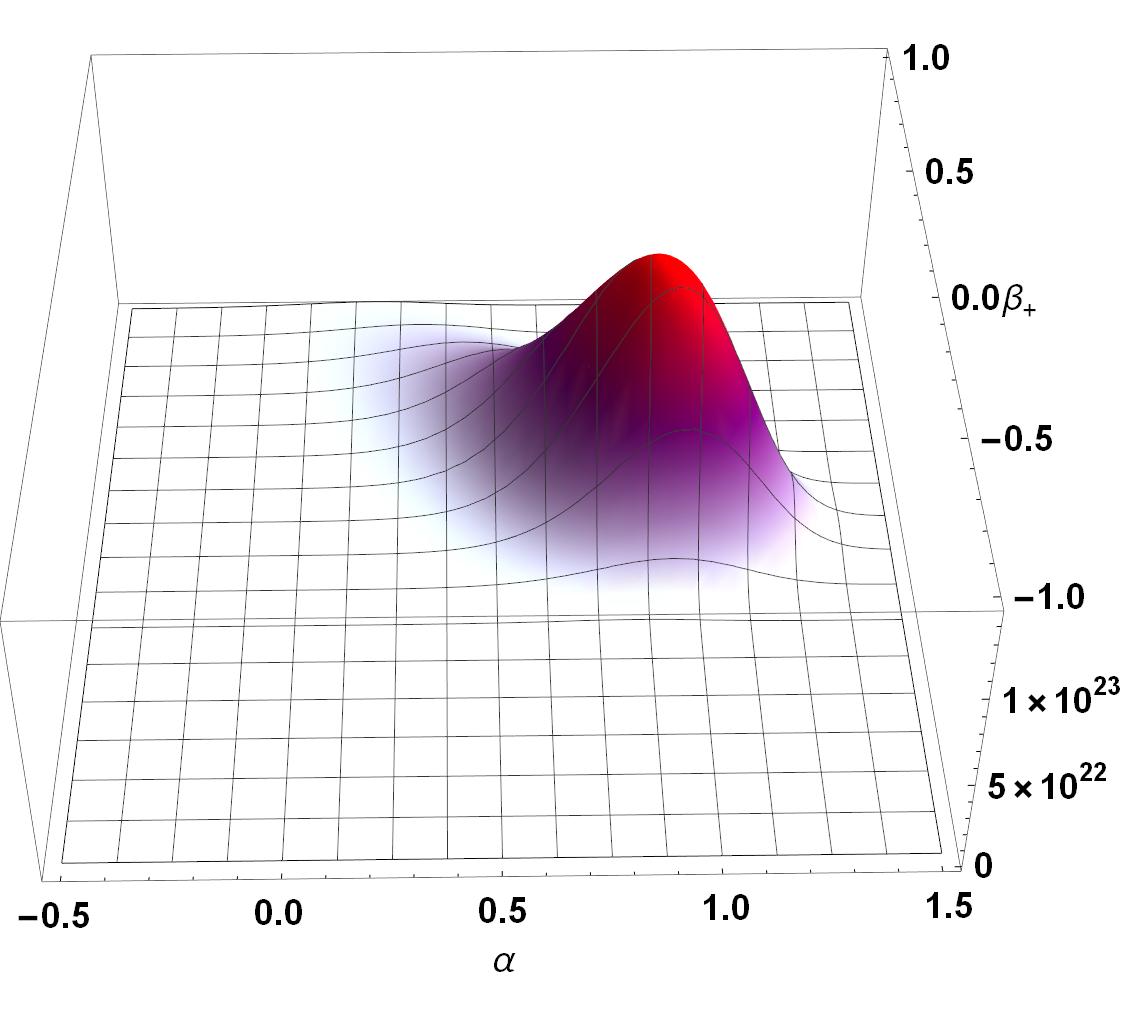}
  \caption{$\phi=0$ \hspace{1mm} $b=2$}
  \label{5b}
\end{subfigure}
\caption{Two plots of (\ref{70}) for two different values for the strength of the electromagnetic field. }
\label{fig5}
\end{figure}
\end{center}
\twocolumngrid\

\begin{equation}\label{62}
\begin{aligned} 
& p_{\alpha} = \frac{\partial \mathcal{S}_{(0)}}{\partial \alpha} \\& p_{+} = \frac{\partial \mathcal{S}_{(0)}}{\partial \beta_{+}} \\& p_{-} = \frac{\partial \mathcal{S}_{(0)}}{\partial \beta_{-}} \\& p_{\phi} = \frac{\partial \mathcal{S}_{(0)}}{\partial \phi}
\end{aligned}
\end{equation}

\begin{equation}\label{63}
\begin{aligned} \dot{\alpha} &=\left.\frac{(6 \pi)^{1 / 2}}{2 e^{3 \alpha}} N\right|_{\text {Eucl }} p_{\alpha} \\ \dot{\beta}_{+} &=\left.\frac{-(6 \pi)^{1 / 2}}{2e^{3 \alpha}} N\right|_{\text {Eucl }} p_{+} \\ \dot{\beta}_{-} &=\left.\frac{-(6 \pi)^{1 / 2}}{2e^{3 \alpha}} N\right|_{\text {Eucl }} p_{-} \\ \dot{\phi} &=\left.\frac{-(6 \pi)^{1 / 2}}{2e^{3 \alpha}} N\right|_{\text {Eucl }} p_{\phi} \end{aligned}
\end{equation}

where $ N|_{\text {Eucl }}$ is the lapse. The lapse $ N|_{\text {Eucl }}$ can be any function of the Misner variables and $\phi$ as long as it never vanishes or changes sign within the range $-\infty$ to $\infty$ of all four variables. To keep our analysis straightforward we will set $ N|_{\text {Eucl }}=\frac{2 e^{\alpha}}{\left(6\pi\right)^{1/2}}$. Using this lapse and (\ref{51}) results in the following classical flow equations  

\begin{equation}\label{64}
\begin{aligned}
&\frac{d \alpha}{d t}=\frac{1}{3} e^{2 \alpha-4 \beta_{+}} \left(2 e^{6 \beta_{+}} \cosh \left(2 \sqrt{3}
   \beta_{-}\right)\pm 1\right) \\&-\frac{\Lambda e^{4 \alpha+4 \beta_{+}}}{9 \pi }-\text{b}^2,
 \end{aligned} 
\end{equation}

\begin{equation}\label{65}
\begin{aligned}
\frac{d \beta_{+}}{d t}=&-2 e^{2 \alpha +2 \beta _+} \cosh \left(2 \sqrt{3} \beta
   _-\right)+\text{bb}^2+\frac{\Lambda e^{4 \alpha +4 \beta _+}}{9 \pi
   }+ \\& \frac{2}{3} e^{2 \alpha -4 \beta _+} \left(2 e^{6 \beta _+} \cosh
   \left(2 \sqrt{3} \beta _-\right)\pm 1\right)),
\end{aligned}
\end{equation}

\begin{equation}\label{66}
\begin{aligned}
\frac{d \beta_{-}}{d t}=-\frac{2 e^{2 \alpha +2 \beta _+} \sinh \left(2 \sqrt{3} \beta
   _-\right)}{\sqrt{3}},
\end{aligned}
\end{equation}

\begin{equation}\label{67}
\begin{aligned}
\frac{d \phi}{d t}=-\frac{\sqrt{\rho}}{2 \sqrt{3}}.
\end{aligned}
\end{equation}
As it can be seen when $\beta_{-}=0$ the time derivative of $\beta_{-}$ vanishes which indicates that if $\beta_{-}$ is initially zero, then it will remain zero indefinitely. Thus the wave functions we will find and analyze correspond to classical trajectories in minisuperspace where $\beta_{-}$ is initially zero. 

Despite this we stress that we are not analyzing the LRS  Bianchi IX models. Even though we will be eventually setting $\beta_{-}=0$ in our calculations, the existence of a $\beta_{-}$ axis will impact our wave functions through the derivatives of $\beta_{-}$ that will appear in our calculations which do not vanish when $\beta_{-}=0$. As the reader can verify, (\ref{51}) is the only $\mathcal{S}_{(0)}$ with matter sources for the Bianchi IX models which possesses this fixed point at $\beta_{-}=0$ in its flow equations. 

If we start with the case when only an aligned electromagnetic field is present($\Lambda=0$) we can insert the entirety of (\ref{51}) into (\ref{42}) which will give us a complicated looking transport equation. However because $\beta_{-}=0$ is a fixed point on the $\beta_{-}$ axis we can set $\beta_{-}=0$ for this complicated transport equation which results from inserting (\ref{51}) into (\ref{42}). The resultant transport equation as the reader can verify can be solved by inserting this ansatz into it

\begin{equation}\label{68}
\begin{aligned}
\mathcal{S}^{1}_{(1) \hspace{1mm \beta{-}=0}}:= x1\alpha +x2\phi,
\end{aligned}
\end{equation}
which yields the following solution 

\begin{equation}\label{69}
\begin{aligned}
\mathcal{S}^{1}_{(1) \hspace{1mm \beta{-}=0}}:=\frac{1}{2} \alpha (-\text{B}-6)-\frac{\sqrt{3} \text{b}^2 \phi}{2 \sqrt{\rho}}.
\end{aligned}
\end{equation}

This is our first quantum correction for our Bianchi IX quantum cosmologies which correspond to classical cosmologies which are formed from a flow in minisuperspace which starts on a fixed point on the $\beta_{-}$ axis. The above quantum correction takes into account the existence of the $\beta_{-}$ axis in the full Bianchi IX models by taking the full derivative of $\mathcal{S}^{1}_{(0 +)}$ in (\ref{42}). An interesting feature of this quantum correction is how the aligned electromagnetic field $b$, stiff matter $\rho$, and the scalar field $\phi$ are coupled. 

Despite it not being the only solution to (\ref{42}) and its simple nature we claim that we are justified in employing it because it generates such fascinating wave functions as we will show shortly. The non-trivial effects that these wave functions imply for the universes that they describe should be chronicled as possible phenomena that a toy model of quantum gravity can induce on the evolution of a quantum universe(s). In addition physically this solution makes sense because the effects of quantum corrections $\mathcal{S}_{( k > 0)}$ in our wave function should become negligible compared to the leading order term  $\mathcal{S}_{( 0)}$ as $\alpha$ increases which is the case as can be seen by comparing (\ref{69}) to  (\ref{51}).

Using this first order quantum correction (\ref{69}) and  

\begin{equation}\label{70}
\begin{aligned}
&\mathcal{S}^{1}_{(0+) \hspace{1mm \beta{-}=0}}:=\frac{1}{6} e^{2 \alpha-4 \beta_{+}} \left(2 e^{6 \beta_{+}}\pm 1\right) \\&+\alpha \text{b}^2+\beta_{+}
   \text{b}^2+\frac{\phi \sqrt{\rho}}{2 \sqrt{3}}.
\end{aligned}
\end{equation}
we can study the quantum cosmology of Bianchi IX 'ground' states which are restricted to the $\beta_{+}$ axis. In addition because our wave functions only possess three minisuperspace variables we can use $\phi$ as our time parameter to obtain wave functions which are easy to interpret via their aesthetic qualities.

Starting with the Bianchi IX models we form the following wave function 

\begin{equation}\label{71}
\begin{aligned}
\psi^{1}_{\hspace{1mm \beta{-}=0}}=e^{-\mathcal{S}^{1}_{(0+) \hspace{1mm \beta{-}=0}}-\mathcal{S}^{1}_{(1) \hspace{1mm \beta{-}=0}}}
\end{aligned}
\end{equation}
and plot them  for two different values $b$ of the aligned electromagnetic field as can be seen in (\ref{fig5}). As can be seen in figure (\ref{5a}) when $b=0$ our wave function is describing a quantum universe which is most likely in a geometric configuration in which its scale factor $\alpha\approx.6$ and $\beta_{+}=0$. However when we turn on our aligned electromagnetic field its wave function is now peaked at larger value of both $\alpha$ and $\beta_{+}$.

\begin{center}
\begin{figure}[h]
\centering
\begin{subfigure}{.4\textwidth}
  \centering
  \includegraphics[scale=.12]{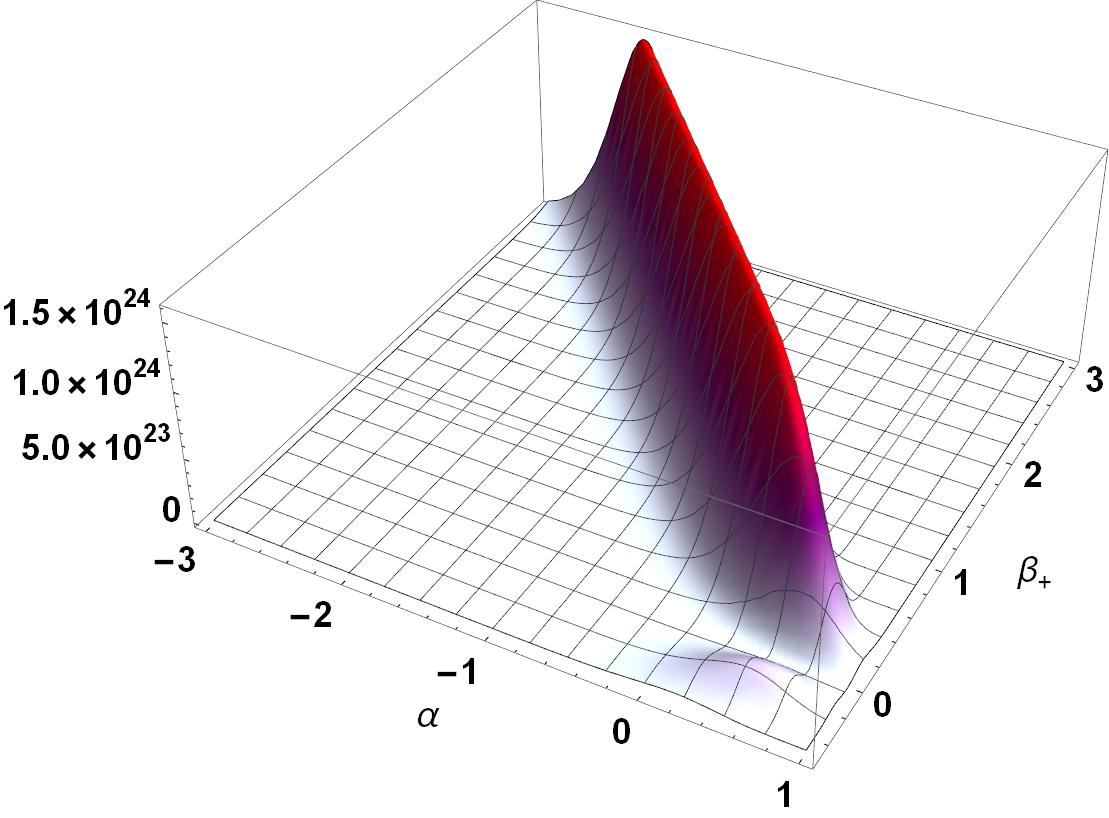}
  \caption{$\phi=0$ \hspace{1mm} $\Lambda=-1$ \hspace{1mm} $b=0$  \hspace{1mm} $n=1$}
  \label{6a}
\end{subfigure}%
\begin{subfigure}{.4\textwidth}
  \centering
  \includegraphics[scale=.12 ]{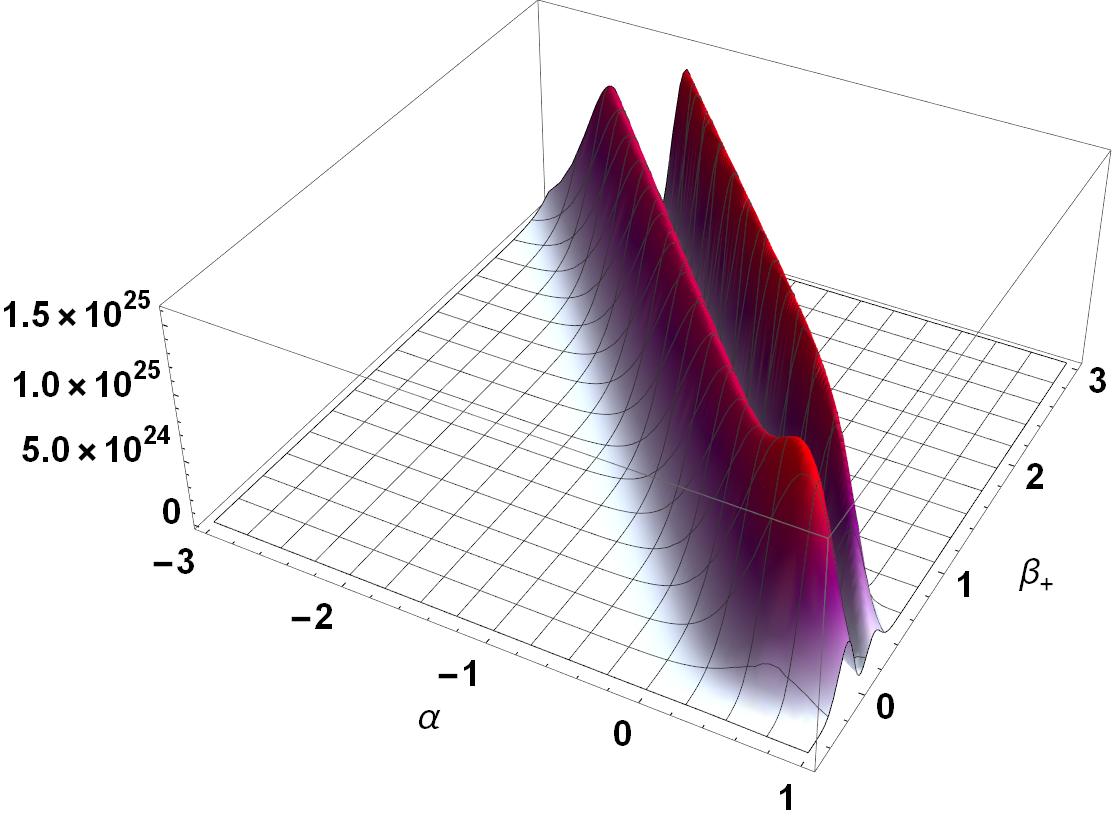}
  \caption{$\phi=0$ \hspace{1mm} $\Lambda=-1$ \hspace{1mm} $b=1.5$ \hspace{1mm} $n=1$}
  \label{6b}
\end{subfigure}
\begin{subfigure}{.4\textwidth}
  \centering
  \includegraphics[scale=.12 ]{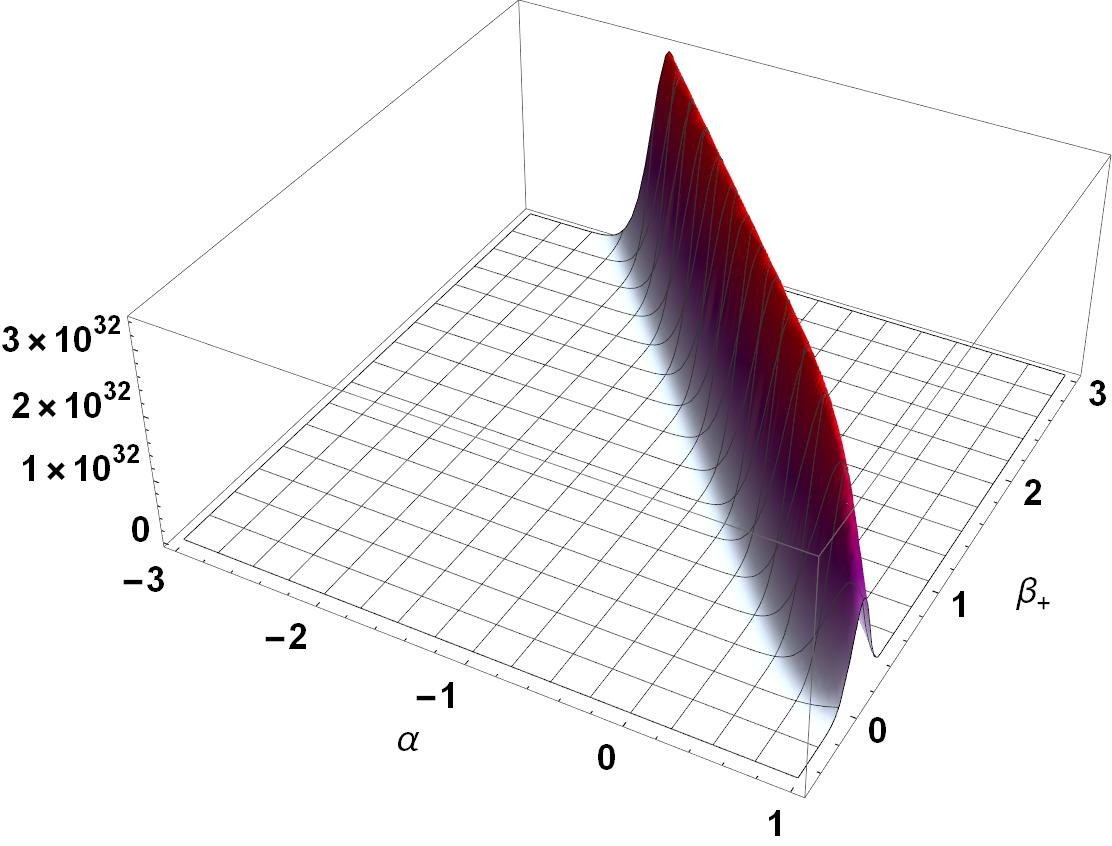}
  \caption{$\phi=0$ \hspace{1mm} $\Lambda=-1$ \hspace{1mm} $b=3$ \hspace{1mm} $n=1$}
  \label{6c}
\end{subfigure}
\caption{Two plots of (\ref{74}) for the first 'excited' states of our Bianchi IX wave functions restricted to the $\beta_{+}$ axis for two different values of the strength of the electromagnetic field $b$.  }
\label{fig6}
\end{figure}
\end{center}

Next we will examine Bianchi IX 'excited' states which are restricted to the $\beta_{+}$ axis when an electromagnetic field and cosmological constant are present. First we insert the entirety of (\ref{51}) into (\ref{48}); then only after taking the derivative of (\ref{51}) with respect to $\beta_{-}$ do we set $\beta_{-}=0$ which results in the following homogeneous transport equation 

\begin{equation}\label{72}
\begin{aligned}
&\text{b}^2 \frac{\partial \Phi_{(0)}}{\partial \alpha}-\text{b}^2 \frac{\partial \Phi_{(0)}}{\partial \beta_{+}}-\frac{1}{3} e^{2
   \alpha -4 \beta _+} \frac{\partial \Phi_{(0)}}{\partial \alpha}-\\&\frac{2}{3} e^{2 \alpha +2 \beta _+}
   \frac{\partial \Phi_{(0)}}{\partial \alpha}  -\frac{2}{3} e^{2 \alpha -4 \beta _+} f\left(\beta
   _+\right)+\frac{2}{3} e^{2 \alpha +2 \beta _+} f\left(\beta _+\right)\\&+2
   \sqrt{3} \sqrt{p} \frac{\partial \Phi_{(0)}}{\partial \phi} +\frac{\Lambda e^{4 \alpha +4 \beta _+}}{9 \pi
   }\frac{\partial \Phi_{(0)}}{\partial \alpha}\\&-\frac{\Lambda e^{4 \alpha +4 \beta _+}}{9 \pi }\frac{\partial \Phi_{(0)}}{\partial \beta_{+}}=0.
\end{aligned}
\end{equation}
The author found the following solutions to this transport equation
\begin{equation}\label{73}
\begin{aligned}
\Phi^{1}_{0 }=&\Bigl(9 \pi  e^{4 \alpha -2 \beta _+}-9 \pi  e^{4 \left(\alpha +\beta
   _+\right)}+27 \pi  \text{b}^2 e^{2 \left(\alpha +\beta _+\right)} \\&+\Lambda e^{6
   \left(\alpha +\beta _+\right)}\Bigr)^{n}
\end{aligned}
\end{equation}
where in order for $\Phi^{1}_{0}$ to be globally defined and smooth we must restrict $n$ to be either a positive integer or zero. Using (\ref{73}) we form the following wave function 

\begin{equation}\label{74}
\begin{aligned}
\Psi^{1}_{\hspace{1mm \beta{-}=0}}=\Phi^{1}_{0+ }e^{-\mathcal{S}^{1}_{(0+) \hspace{1mm \beta{-}=0}}-\mathcal{S}^{1}_{(1) \hspace{1mm \beta{-}=0}}}
\end{aligned}
\end{equation}
which we plot in (\ref{fig6}). Figure 6 emphatically shows that when both a negative cosmological constant and an electromagnetic field are present the effects of the electromagnetic field are dependent upon its strength. When the field is relatively weak as can be seen (\ref{6a}) and (\ref{6b}) it can cause an additional highly probably geometric state to come into existence which our quantum universe can tunnel in and out of. However when the electromagnetic field is strong it can also destroy a highly probable state, leaving only one highly probable state that a quantum universe can be in. This ability to create an additional state that a quantum universe can tunnel in and out of is similar to how non-commutativity in the minisuperspace variables can cause new quantum states to emerge in the quantum Kantowski-Sachs\cite{garcia2002noncommutative} and Bianchi I models\cite{vakili2007bianchi}. Chronicling these effects of our aligned electromagnetic field can help us understand what our early universe could have been like. 

\onecolumngrid\
\twocolumngrid\

\section{\label{sec:level7}Quantum Taub Models With A Scalar Field}

\onecolumngrid\
\begin{center}
\begin{figure}[h]
\centering
\begin{subfigure}{.4\textwidth}
  \centering
  \includegraphics[scale=.15 ]{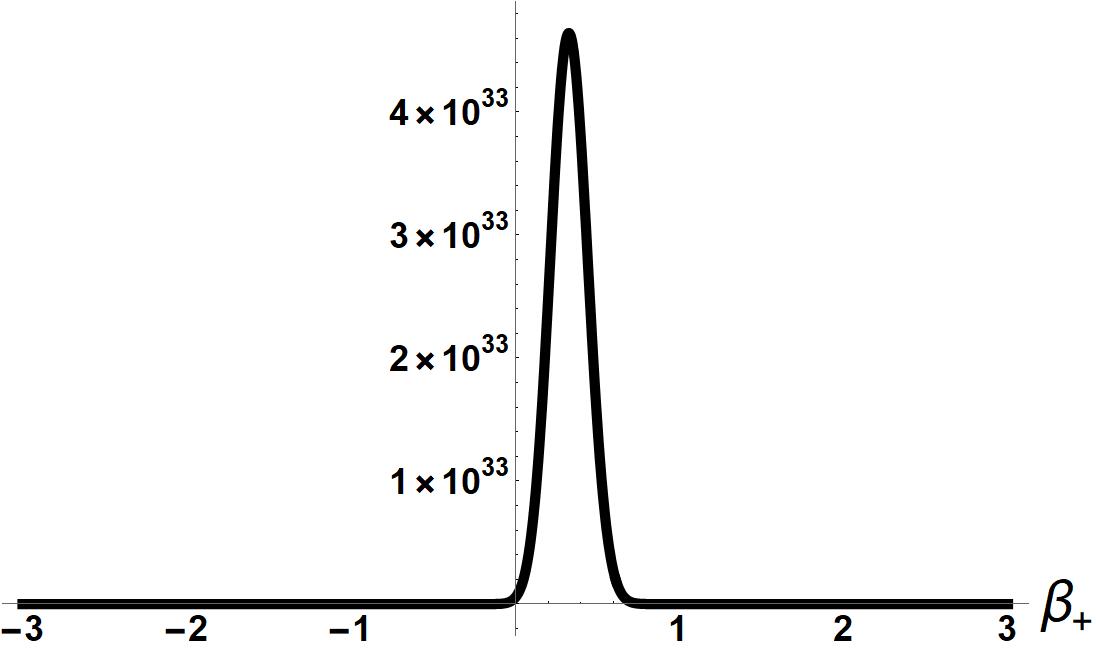}
  \caption{ $\alpha=1$ \hspace{1mm}  $c=-1$ \hspace{1mm} $\phi=0$}
  \label{7a}
\end{subfigure}%
\begin{subfigure}{.4\textwidth}
  \centering
  \includegraphics[scale=.15 ]{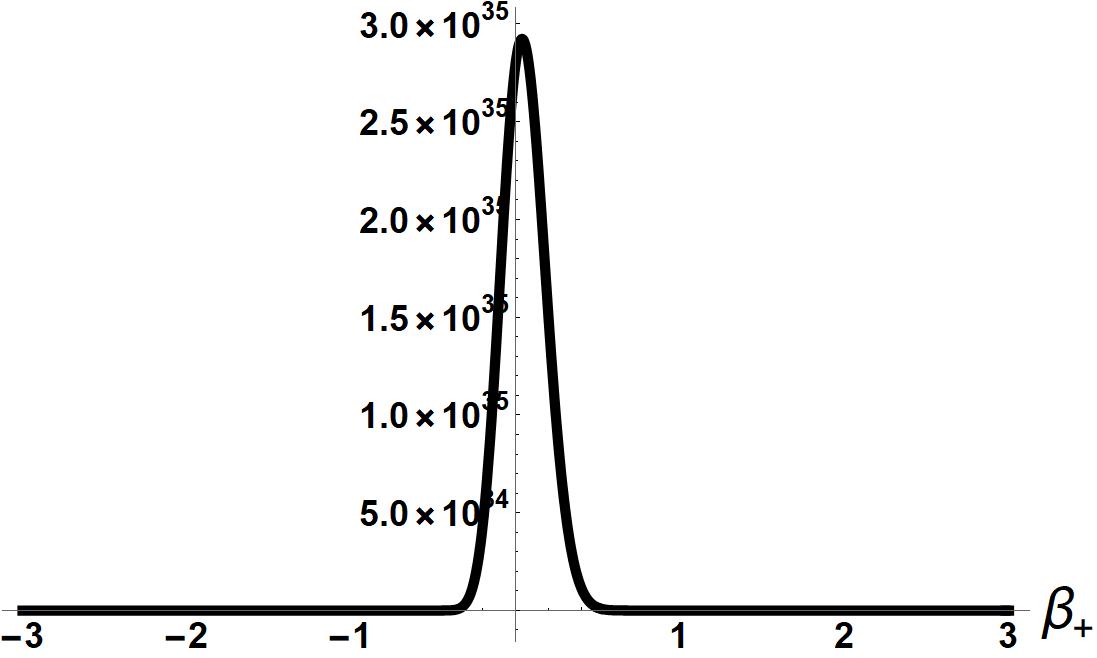}
  \caption{ $\alpha=1$ \hspace{1mm}  $c=-1$ \hspace{1mm} $\phi=-5$}
  \label{7b}
\end{subfigure}
\begin{subfigure}{.4\textwidth}
  \centering
  \includegraphics[scale=.15]{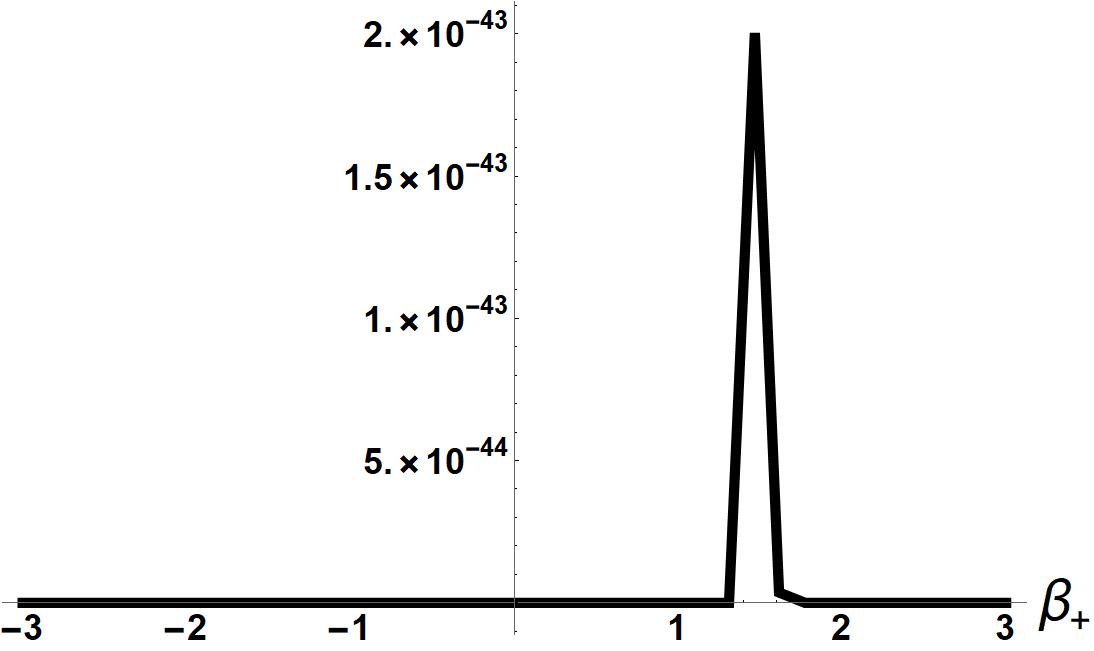}
  \caption{ $\alpha=1$ \hspace{1mm}  $c=-1$ \hspace{1mm} $\phi=5$}
  \label{7c}
\end{subfigure}
\caption{These are three plots of our Taub wave functions when an exponential scalar potential is present.} 
\label{fig7}
\end{figure}
\end{center}
\twocolumngrid\

The author found the following $\mathcal{S}_{(0)}$ for the quantum Taub models when an exponential scalar potential and an aligned electromagnetic field are present

\begin{equation}\label{75}
\begin{aligned}
&\mathcal{S}^{4}_{(0)}:=\frac{1}{6} e^{2 \alpha-4 \beta_{+}}+\frac{1}{3} e^{2 \alpha+2 \beta_{+}}-\frac{c}{8}  e^{4 \alpha-2 \beta_{+}+\phi} \\&-\alpha \text{b}^2+\frac{\beta_{+} \text{b}^2}{2}-\frac{\text{b}^2 \phi}{4},
\end{aligned}
\end{equation}
where c can in principle be any real number. However for the purposes of elucidating the points the author wishes we to make we will assume $c<0$. Utilizing (\ref{42}) the author found the following $\mathcal{S}_{(1)}$ quantum correction to (\ref{75})

\begin{equation}\label{76}
\begin{aligned}
\mathcal{S}^{4}_{(1)}:=\frac{1}{2} \alpha (-\text{B}-2)-\beta_{+}-\frac{\phi}{2}.
\end{aligned}
\end{equation}
The existence of (\ref{75}) and (\ref{76}),  suggest that an analogous closed form solution involving an exponential scalar field potential may exist for the Bianchi IX models as well. 

First we will plot $e^{-\mathcal{S}^{4}_{(0)}-\mathcal{S}^{4}_{(1)}}$ when $B=0$ and $b=0$ and use both $\alpha$ and $\phi$ as our internal clocks.

Our plots(\ref{fig7}) indicate that as our scalar potential decays the universes described by these wave functions approache isotropy. However when our scalar potential increases in magnitude these universes become more anisotropic. Similar qualitative behavior may be present for other types of scalar potentials such as $\phi^{n}$.

These effects are interesting in themselves, however to further determine how typical or robust they are, more general scalar fields need to be studied. To accomplish this in the future various approximate techniques which do not rely on obtaining closed form solutions would need to be employed. However for the meantime these findings encourage further work to be done to determine how scalar fields effect quantum cosmological evolution in anisotropic models. 

\begin{center}
\begin{figure}[h]
\centering
\begin{subfigure}{.4\textwidth}
  \centering
  \includegraphics[scale=.15 ]{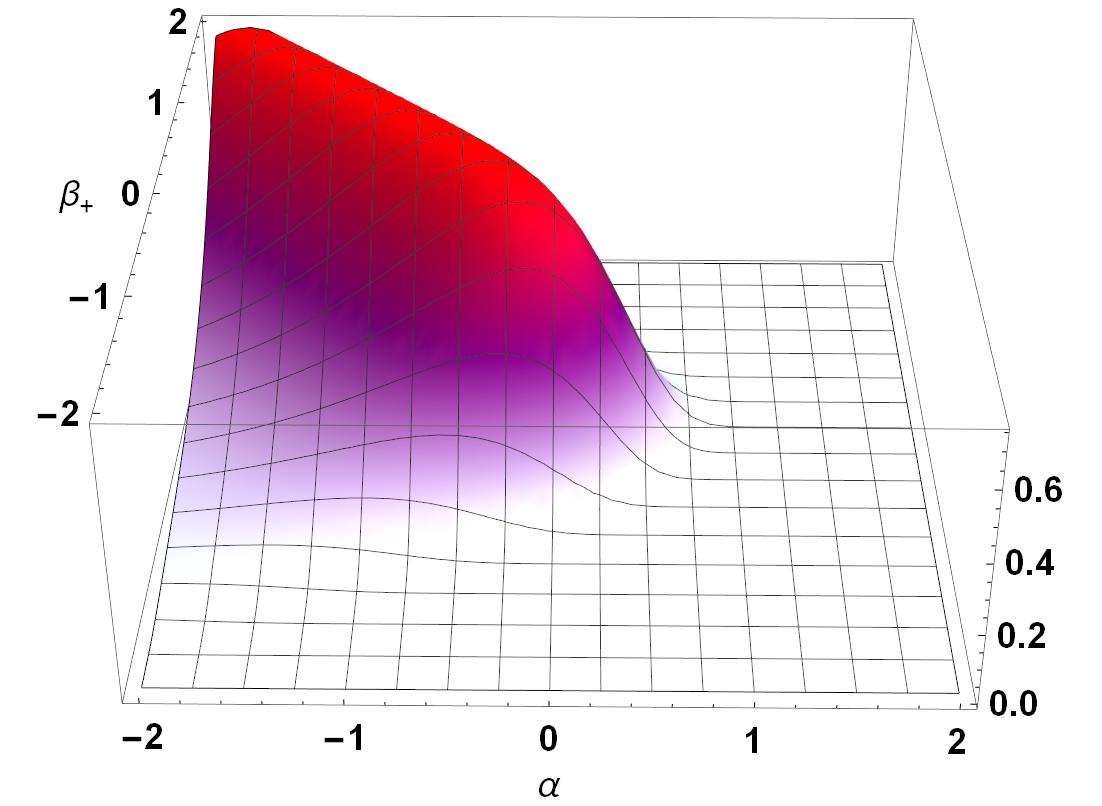}
  \caption{ $b=0$ \hspace{1mm}  $c=-1$ \hspace{1mm} $\phi=0$}
  \label{8a}
\end{subfigure}%
\begin{subfigure}{.4\textwidth}
  \centering
  \includegraphics[scale=.15 ]{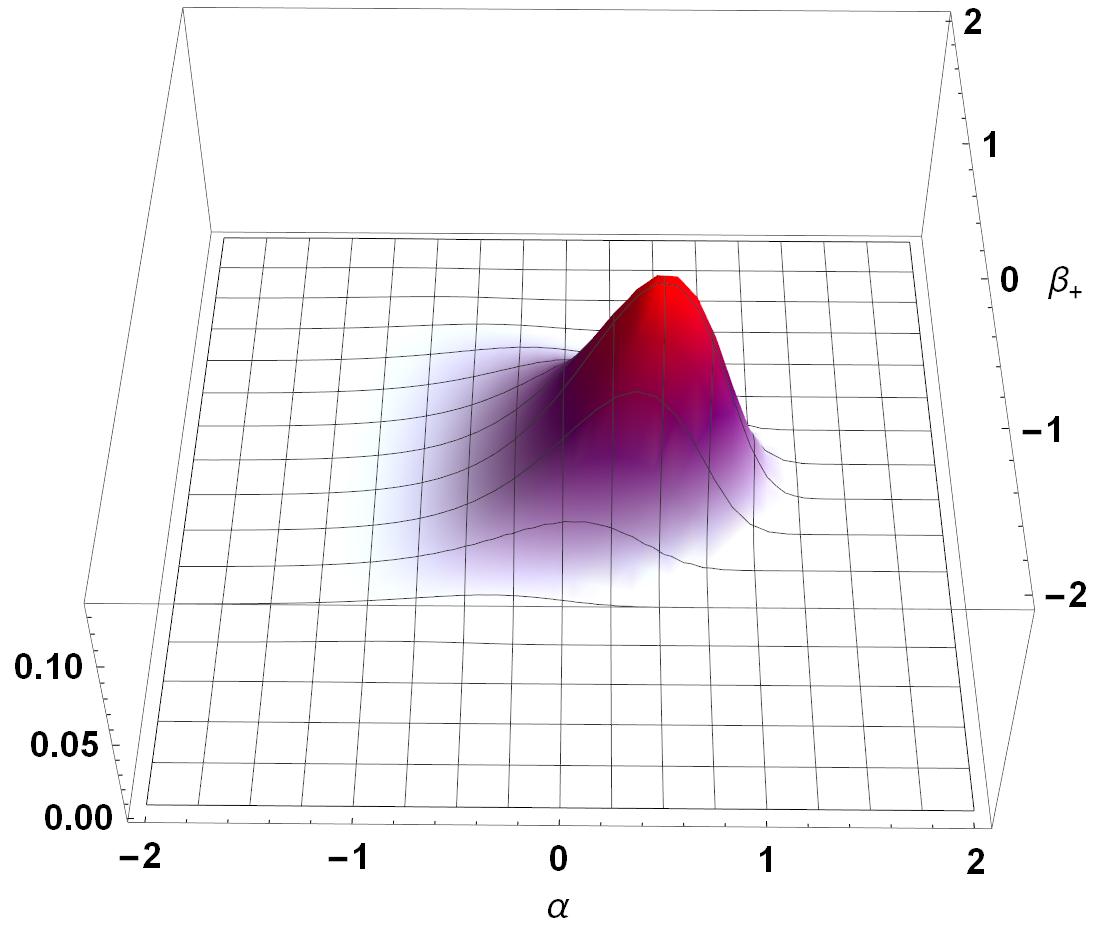}
  \caption{ $b=1.5$ \hspace{1mm}  $c=-1$ \hspace{1mm} $\phi=-2$}
  \label{8b}
\end{subfigure}
\begin{subfigure}{.4\textwidth}
  \centering
  \includegraphics[scale=.15]{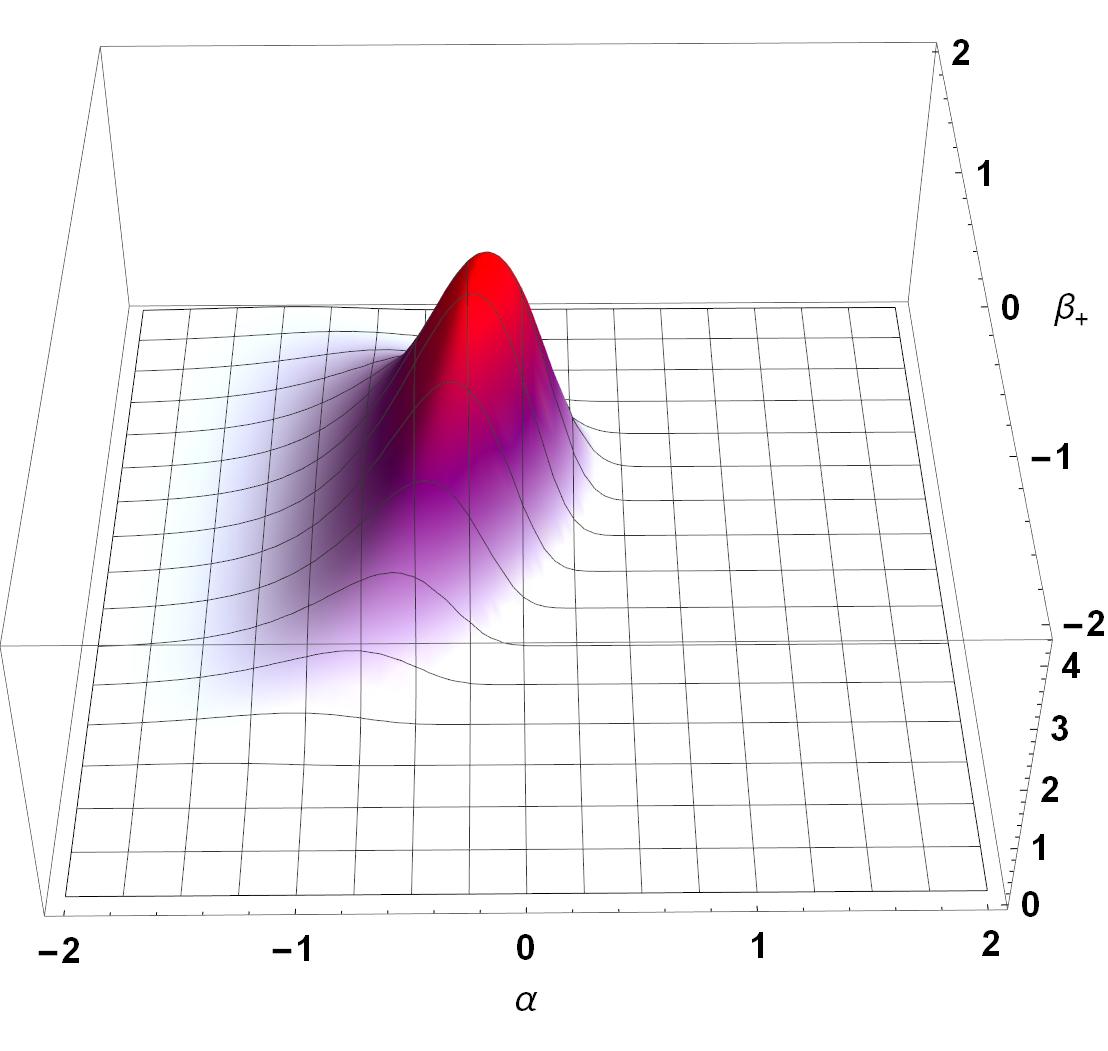}
  \caption{ $b=1.5$ \hspace{1mm}  $c=-1$ \hspace{1mm} $\phi=3$}
  \label{8c}
\end{subfigure}
\caption{These are three plots of our Taub wave functions when an exponential scalar potential and aligned electromagnetic are present.} 
\label{fig8}
\end{figure}
\end{center}

For the case when an aligned electromagnetic field is present we will use $\phi$ as our sole internal clock. As can be seen in figure 8, if we use $\phi$ as our internal clock, the aligned electromagnetic field induces some dramatic effects on our wave functions. If we compare (\ref{8a}) to (\ref{8b}) we see that the electromagnetic field causes the very 'fuzzy' wave function in (\ref{8a}) to become sharp, or peaked at a certain value $\alpha$ and $\beta_{+}$. This effect would be present even if we kept $\phi=0$. As $\phi$ increases, our wave functions travel in the negative $\alpha$ direction which indicates that the universes we are describing are more likely to be smaller in size as $\phi$ grows. In addition as $\phi$ grows it moves in the positive $\beta_{+}$ direction which indicates its anisotropy in general increases as $\phi$ increases. The fact that the Euclidean-signature semi classical method was able to generate such a wealth of information for the solutions of (\ref{8}) further shows its ability to prove\cite{bae2015mixmaster,moncrief2014euclidean} the existence of solutions to Lorentzian signature equations. 

\onecolumngrid\
\twocolumngrid\

\section{\label{sec:level8}Closed Form Solutions To Bianchi IX and VIII WDW}
If we start with (\ref{51}) and set $\Lambda=0$, and solve the  equation which results from inserting (\ref{51}) into (\ref{42}), we obtain the following $\mathcal{S}_{(1)}$ quantum correction

\begin{equation}\label{77}
\begin{aligned}
& \mathcal{S}^{1}_{(1 \pm)}:=\alpha \text{x1}+\frac{1}{8} (\text{B}+2 \text{x1}+6) \log \left(\sinh \left(2
   \sqrt{3} \beta_{-}\right)\right) \\&+\beta_{+} \left(\frac{1}{4}
   (-\text{B}-6)-\frac{\text{x1}}{2}\right) \\& \mp\phi \left(\frac{\sqrt{3}
   \text{b}^2 (\text{B}+2)}{8 \sqrt{\rho}} + \frac{\sqrt{3} \text{b}^2
   \text{x1}}{4 \sqrt{\rho}}\right),
\end{aligned}
\end{equation}
where x1 is an arbitrary constant, and B is the Hartle-Hawking ordering parameter. 
Technically this $\mathcal{S}_{(1)}$ term is not smooth and globally defined, however we can form smooth and globally defined wave functions from it if we restrict $(\text{B}+2 \text{x1}+6) $ to be $-8y$, where y is either zero or a positive integer. Using (\ref{43}) we can obtain solutions to (\ref{7}) of the form  $e^{- \mathcal{S}^{1}_{(0 \pm)} -\mathcal{S}^{1}_{(1 \pm)}}$, when $\Lambda=0$ and our aligned electromagnetic field is given by  $2b^{2}e^{2\alpha-4\beta_{+}}$ if the constants $(x1,\rho,b,B)$ in (\ref{77}) satisfy the following relations 

\begin{equation}\label{78}
\begin{aligned}
& \frac{3 \text{b}^4 (\text{B}-2 \text{x1}+2)^2}{32 \rho}-2 \text{B}
   \text{x1}+2 \text{x1}^2=0 \\& \text{B}+2 \text{x1}+6=0.
\end{aligned}
\end{equation}
Beyond these closed form solutions we can also construct smooth and globally defined wave functions using (\ref{51}) and (\ref{77}) as long as $(\text{B}+2 \text{x1}+6) $ is a negative integer which is a multiple of 8. The question of solving the higher order transport equations for the Bianchi IX and VIII models when an aligned electromagnetic field is present in the other two perpendicular directions could be the topic of a future investigation. For now we will be content with this closed form solution for an aligned electromagnetic field.

For the sake of completeness we will report a few other closed form solutions that the author found for the vacuum Bianchi IX and VIII models which were reported in earlier works. For the vacuum Bianchi IX models when $\rho=0$, $\Lambda=0$, and $b=0$, Joseph Bae\cite{bae2015mixmaster} found two families of quantities which are conserved under the flow produced by (\ref{51}), and thus satisfy (\ref{48}). The author then modified those quantities so that they are conserved under the flow given by (\ref{51}) for the Bianchi VIII models. Thus we will use the following quantities which satisfy (\ref{48}) for the vacuum Bianchi IX and VIII models

\begin{equation}\label{79}
\begin{aligned}
&\Phi^{1}_{0 \pm}:=S^{m1}O^{m2}, \\& S= \left(e^{4 \alpha-2 \beta_{+}} \left(e^{6 \beta_{+}}\mp\cosh \left(2 \sqrt{3}
   \beta_{-}\right)\right)\right), \\& O=\left(e^{4 \alpha-2 \beta_{+}} \sinh \left(2
   \sqrt{3} \beta_{-}\right)\right),
\end{aligned}
\end{equation}
where m1 is either zero or a positive integer for the Bianchi IX case, or can be any number for the Bianchi VIII models, m2 is always either zero or a positive integer. Joseph Bae found the following solution to (\ref{42}) for the vacuum Bianchi IX models, which also happens to satisfy (\ref{42}) for the vacuum Bianchi VIII models  

\begin{equation}\label{80}
\begin{aligned}
\mathcal{S}_{(1 \hspace{1mm} Bae)}:=-\frac{1}{2} \alpha (\text{B}+6).
\end{aligned}
\end{equation}
Solutions of the form $\Phi^{1}_{0 \pm}e^{- \mathcal{S}_{(0 \pm)} -\mathcal{S}^{1}_{(1 \hspace{1mm} Bae)}}$ exist for the vacuum Bianchi VIII and IX models for the following values of $(B,m1,m2)$

\begin{equation}\label{81}
\begin{aligned}
&(B=\pm6, m1=0, m2=0) \\&(B=\pm 2\sqrt{33}, m1=1, m2=0) \\& (B=\pm 2\sqrt{33}, m1=0, m2=1)
\end{aligned}
\end{equation}
More information on these closed form solutions can be found in\cite{berkowitz2019new}.

\section{\label{sec:level9}Concluding Remarks}
Using the Euclidean-signature semi classical method we have obtained in closed form some very interesting wave functions of the universe for the Bianchi IX, VIII and Taub models. By doing so we have shown how certain matter such as a negative cosmological constant, an aligned electromagnetic field, and scalar field can effect the evolution of a quantum universe. Most notably we have provided additional evidence\cite{berkowitz2020towards,berkowitz2020atowards} that an aligned electromagnetic field can modify the states/peaks that are present in 'excited' states. 

The next step in establishing how an electric/magnetic field can effect the 'excited' states of the WDW equation for Bianchi A models would be to study the Bianchi I models with a magnetic field and cosmological constant. Afterwards it would be instructive to investigate how general/non-aligned electric/magnetic fields with inhomogeneous perturbations can effect the evolution of quantum universes. 

Nonetheless the toy model we studied yielded interesting wave functions whose effects deserve to be chronicled as possible phenomena that quantum gravity can induce on the evolution of a universe. In addition by obtaining these wave functions for the Lorentzian signature WDW equation by finding closed form solutions to the Euclidean signature Hamilton-Jacobi equation we have further shown the utility of the Euclidean-signature semi classical method.

The author hopes that the plethora of results obtained in this paper, in conjunction with all of the results previously obtained by this method will hope promote its use in solving important problems in both field theory, such as the 'mass gap' problem and in quantum cosmology. An attractive feature of the Euclidean-signature semi classical method as it applies to interacting bosonic field theories is that it doesn't require splitting the theory up into one part which is linear(non-interacting) and another part which is a nonlinear(interacting) perturbation. This allows the full interacting nature of the field theory to be present\cite{marini2019euclidean,marini2020euclidean} at every level of its analysis.

\section{\label{sec:level11}ACKNOWLEDGMENTS}
 
I am grateful to Professor Vincent Moncrief for valuable discussions at every stage of this work. I would also like to thank George Fleming for facilitating my ongoing research in quantum cosmology. Daniel Berkowitz acknowledges support from the United States Department of Energy through grant number DE-SC0019061. I also must thank my aforementioned parents.

\bibliography{ref}

\end{document}